\documentclass[11pt]{article}
\usepackage[a4paper, total={6in, 10in}]{geometry}
\usepackage{mathrsfs,theorem,amssymb,amsmath,bm,authblk}
\usepackage{graphicx}
\usepackage[bookmarks=true,colorlinks=true,urlcolor=blue,linkcolor=blue,citecolor=blue]{hyperref}

\newtheorem{theorem}{Theorem}
\newtheorem{corollary}{Corollary}
\newtheorem{definition}{Definition}
\newtheorem{lemma}{Lemma}

\newcommand{\hil}{\mathcal{H}}
\newcommand{\R}{\mathcal{R}}
\newcommand{\braket}[2]{\langle #1 | #2 \rangle}
\newcommand{\ketbra}[1][]{| #1 \rangle \langle #1 |}
\newcommand{\ket}[1][]{| #1 \rangle}
\newcommand{\bra}[1][]{\langle #1 |}
\newcommand{\vecn}{{\bm n}}
\newcommand{\trace}{\mathrm{Tr}}
\newcommand{\id}{\mathbb{I}}
\newcommand{\real}{\mathbb{R}}
\newcommand{\antiu}{\Theta}
\newcommand{\conj}{\theta}
\newcommand{\compt}{\zeta}
\newcommand{\magic}{\mu}

\newcommand{\rpart}{\mathrm{Re}}
\newcommand{\ipart}{\mathrm{Im}}
\newcommand{\SO}{\mathbb{SO}}
\newcommand{\SU}{\mathbb{SU}}
\newcommand{\swp}{\mathrm{SWAP}}
\newcommand{\qed}{\hfill\blacksquare}

\begin{document}

\title{Non-locality of conjugation symmetry: characterization and examples in quantum network sensing}

\author[1]{Jisho Miyazaki\thanks{23v00005@gst.ritsumei.ac.jp}}
\affil[1]{Ritsumeikan University BKC Research Organization of Social Sciences, 1-1-1 Noji-higashi, Kusatsu, Shiga, 525-8577 Japan}
\author[2]{Seiseki Akibue\thanks{seiseki.akibue@ntt.com}}
\affil[2]{NTT Communication Science Laboratories, 3-1 Morinosato-Wakamiya, Atsugi-shi, Kanagawa, 243-0198
Japan}

\date{\today}

\maketitle

\begin{abstract}
Some quantum information processing protocols necessitate quantum operations that are invariant under complex conjugation. In this study, we analyze the non-local resources necessary for implementing conjugation-symmetric measurements on multipartite quantum networks. We derive conditions under which a given multipartite conjugation can have locally implementable symmetric measurements. In particular, a family of numbers called the ``magic-basis spectrum'' comprehensively characterizes the local measurability of a given 2-qubit conjugation, as well as any other properties that are invariant under local unitary transformations. We also explore the non-local resources required for optimal measurements on known quantum sensor networks by using their conjugation symmetries as a guide.
\end{abstract}

\tableofcontents

\section{Introduction}
Quantum theory formalized on real Hilbert spaces has been a longstanding subject of research. The original idea traces back to St\"{u}eckelberg's work in the 1960s \cite{stueckelbergQuantumTheoryReal1960}. Since then, researchers have recognized several disparities between real and complex quantum theories.

``Real'' quantum systems are not only purely mathematical abstractions; they manifest naturally in various quantum information processing scenarios. Wootters demonstrated that constraining quantum systems to the real domain facilitates optimal information transfer \cite{woottersOptimalInformationTransfer2013}. It is possible to embed complex quantum states into a real system for simulating unphysical transformations \cite{casanovaQuantumSimulationMajorana2011,zhangTimeReversalCharge2015} and for calculating entanglement measures efficiently \cite{dicandiaEmbeddingsimulator2013,chenEfficientMeasurement2016}. Of particular relevance to this paper is \cite{miyazaki2022imaginarityfree}, in which it is observed that diverse metrology schemes, including the renowned phase estimation \cite{leekokrossetastone2002,giovannettiQuantumMetrology2006,giovannettiAdvancesQuantumMetrology2011}, exclusively rely on real subspaces within the complex Hilbert space. Consequently, these schemes have been termed ``imaginarity-free'' estimations.

In the realm of real quantum theory, composite systems are defined as tensor products of real subsystems, thereby ensuring an equitable comparison with their complex counterparts. This perspective entails considering a specific real subspace within multipartite complex Hilbert spaces. The impact of restricting multipartite systems to real scenarios has been observed in state discrimination \cite{wootters1990local,chiribellaProbabilisticTheory2010,hardyLimitedHolismRealVectorSpace2012,chiribellaProcessTomography2021,wuOperationalResourceTheory2021,wuResourceTheoryImaginarity2021}, monogamy of entanglement \cite{woottersEntanglementSharingRealVectorSpace2012,woottersRebitThreetangleIts2014}, non-locality \cite{renouQuantumTheoryBased2021a,bednorz2022optimaldiscriminationreal}, and most recently, in field-dependent entanglement \cite{chiribellaPositiveMapsEntanglement2023}. However, it is important to note that, in general, a real subspace of a composite complex space cannot be represented as a mere composition of real subspaces. For example, in a two-qubit system, the real span of the magic basis \cite{bennettMixedStateEntanglement1996,hillEntanglementPairQuantum1997a,woottersEntanglementFormationArbitrary1998} consists solely of maximally entangled vectors. Through leveraging the mathematical tools developed in the context of real quantum theory, the investigation of novel real subspaces is expected to expand the prospects and enhance the efficacy of quantum information processing.

In particular, the non-local structure of general real subspaces, as exemplified by the magic basis, has remained unexplored until now, despite its significance in practical information processing scenarios. Here, the term ``non-local structure'' denotes the distribution of entanglement within the real subspace, with precise definitions to be provided subsequently. Indeed, when adapting imaginarity-free estimation \cite{miyazaki2022imaginarityfree} to a network of distributed sensors, comprehending the non-local structure of the real subspace becomes pivotal in determining the consumption of communication resources.

In this article, we analyze the non-local structure of real subspaces through the lens of associated conjugations. Conjugations are specific antiunitary operators on complex Hilbert spaces, and they are in one-to-one correspondence with the real subspaces, which are symmetric under their action. Upon this correspondence reviewed in section \ref{subsec:antiunitary-operators}, investigating real subspaces is tantamount to studying conjugations. Although conjugation is not a physical operation, its mathematical structure resembles that of unitary operators. We hence explore the non-local structure of conjugations by exploiting the non-locality of unitary operators.

First, we examine the local unitary (LU) equivalence classes of multipartite conjugations. By identifying distinct operators that differ only locally, their non-local properties become more apparent. We define LU-equivalence classes of general multipartite conjugations. Characterizing the classes is, however, difficult in general. The same applies to LU-equivalence classes of unitary gates, where clear solution has been known only for two-qubit gates through the Kraus-Cirac decomposition \cite{krauscirac2001}. We employ similar techniques to obtain a canonical decomposition of two-qubit conjugations and achieve a complete characterization of their LU-equivalence classes.

Second, we characterize the non-locality of a real subspace through measurements symmetric under the associated conjugation. We focus on the conjugation-symmetric rank-$1$ positive-operator-valued measures (POVMs). In order to describe the communication resources required for their distributed implementation, we classify the POVMs as product, separable, or entangled measurement. We present several conditions for a conjugation to have symmetric product measurements. Conjugations associated with product real systems are not the only examples of this kind. By using the conditions, we identify conjugations with symmetric nonproduct separable measurements.

When restricted to two-qubit systems, conjugation-symmetric measurements can be characterized by canonical decomposition of two-qubit conjugations. We show that if a rank-$1$ POVM is symmetric under a conjugation $\conj$ and implementable by local operations and classical communication, we can find a $\conj$-symmetric and rank-1 POVM implementable only by non-adaptive local operations. Additionally, for given two-qubit conjugation, we identify a conjugation-symmetric rank-$1$ POVM that minimizes the average entanglement of POVM elements and obtain a closed formula for the minimum value. The minimum average entanglement is a lower bound of the state-entanglement required for implementing the measurement.

Conjugation-symmetric measurements play a pivotal role in imaginarity-free estimation \cite{miyazaki2022imaginarityfree}, a method of quantum parameter estimation. In particular, implementing an optimal measurement with low entanglement consumption is desirable when estimating parameters encoded in a network of quantum sensors. Zhou et al.\cite{zhouSaturatingQuantumCramer2020} demonstrated that optimal LOCC measurements always exist for any single-parameter estimation from multipartite pure states. However, finding optimal measurements for multiparameter estimations typically requires numerical calculations \cite{albarelliHolevoBound2019,sidhuTightBounds2021}, and, as yet, there is no algorithm to simultaneously optimize precision and entanglement consumption. For this reason, surveys on optimal measurements for multiparameter estimation on sensor networks have been restricted to particular examples \cite{giovannettiQuantumMetrology2006,boixoQuantumlimitedMetrologyProduct2008,royExponentiallyEnhancedQuantum2008,ramsLimitsCriticalityBasedQuantum2018,eldredgeOptimalSecureMeasurement2018,proctorMultiparameterEstimationNetworked2018}. Our analysis on the non-locality of real subspaces makes it possible to determine the existence of an optimal product measurement for a given imaginarity-free estimation. Here, measurements are referred to be optimal when they saturate the quantum Cram\'{e}r-Rao bound (QCRB) \cite{helstromMinimumVarianceEstimates1968,helstrom1976}. Finally, we provide additional insights into how conjugation-symmetric measurements can be fully utilized in imaginarity-free estimations on sensor networks.

\section{Background}\label{sec:preliminary}
Let us introduce the notation used throughout this paper. We use capital alphabetical characters $A,~M,~U,~V, \ldots$ to denote matrices. Linear operators are denoted by capital letters with circumflexes $\hat{U},~\hat{V}, \ldots$. Bases of Hilbert spaces refer to orthonormal ones and are denoted by small Greek letters $\psi,\eta,\compt,\magic,\ldots$. Among the bases, $\compt$ denotes the computational basis of any Hilbert space in question, and $\magic = \{ \ket[\mu_1] ,~ \ket[\mu_2] ,~ \ket[\mu_3] ,~ \ket[\mu_4] \}$ denotes the magic basis of a two-qubit system \cite{bennettMixedStateEntanglement1996} defined by
\begin{equation}
 \label{magic} \ket[\magic_1] = \ket[\Psi_+],~ \ket[\magic_2] = i \ket[\Psi_-],~ \ket[\magic_3] = i \ket[\Phi_+],~ \ket[\magic_4] = \ket[\Phi_-],
\end{equation}
where $\ket[\Psi_\pm]$ and $\ket[\Phi_\pm]$ are Bell-basis vectors,
\begin{equation}
	\label{Bell}	\ket[\Psi_\pm] := \frac{\ket[00] \pm \ket[11]}{\sqrt{2}}, ~ \ket[\Phi_\pm] := \frac{\ket[01] \pm \ket[10]}{\sqrt{2}}.
\end{equation}
The matrix representation of a linear operator $\hat{U}$ in a basis $\psi$ is denoted by $[\hat{U}]^\psi$.

\subsection{Antiunitary operators}\label{subsec:antiunitary-operators}
An operator $\antiu:\hil \rightarrow \hil$ on a Hilbert space $\hil$ is said to be antilinear if it satisfies $\antiu (c_1 \ket[\psi_1] + c_2 \ket[\psi_2]) = c_1^\ast \antiu \ket[\psi_1] + c_2^\ast \antiu \ket[\psi_2]$ for any pair of vectors $\ket[\psi_1]$, $\ket[\psi_2]$ and any pair of complex numbers $c_1$, $c_2$. The Hermitian adjoint of an antilinear operator is defined by $\bra[\psi_1] ( \antiu^\dagger \ket[\psi_2]) = \bra[\psi_2] ( \antiu \ket[\psi_1])$. If $\antiu^\dagger$ equals the inverse $\antiu^{-1}$, $\antiu$ is called antiunitary. Hermitian (in the sense of $\antiu^\dagger = \antiu$) antiunitary operators are called conjugations. In this article, antilinear operators are represented by capital theta $\antiu$, while small theta $\conj$ is used for stressing that the operator is a conjugation. We consider finite-dimensional Hilbert spaces in this article.

A real Hilbert subspace of complex Hilbert space $\hil$ refers to a subset of $\hil$ that is closed under summation and multiplication by real numbers, and that $\langle \psi_1 | \psi_2 \rangle = \langle \psi_2 | \psi_1 \rangle$ holds for any pair of its elements $\ket[\psi_1]$ and $\ket[\psi_2]$. A real Hilbert subspace $\R$ of $\hil$ is said to be maximal if the real dimension of $\R$ is equal to $\dim \hil$. We henceforth call maximal real Hilbert subspaces as ``real subspaces'' for simplicity.

There are continuously many bases in a single real subspace.
If $\{ \ket[\psi_j] \}_{j=1,...,\dim \hil}$ is a basis of $\R$, so is the basis $\{ R \ket[\psi_j] \}_{j=1,...,\dim \hil}$, where $R$ is a real orthogonal matrix. For example, the real-spans of two different bases of a two-qubit system,
\begin{align}
 \label{basis2} &\left\{ \ket[00],~\ket[01],~\ket[10],~\ket[11] \right\},\\
 \label{basis1} &\left\{ \ket[\Psi_+],~ \ket[\Psi_-],~ \ket[\Phi_+],~ \ket[\Phi_-] \right\},
\end{align}
where the Bell basis \eqref{basis1} is defined by \eqref{Bell}, result in the same real subspace.

A conjugation on finite dimensional Hilbert space $\hil$ defines a unique real subspace of $\hil$ which is symmetric under its action, and all real subspaces of $\hil$ is obtained in this way. From conjugation $\conj$ define the real subspace $\R_\conj$ by
\begin{equation}
	\label{conjtoreal}	\R_\conj := \left\{ \ket[\psi] \in \hil ~\middle|~ \conj \ket[\psi]= \ket[\psi] \right\} = \left\{ \ket[\psi] + \conj \ket[\psi]  ~\middle|~ \ket[\psi] \in \hil \right\}.
\end{equation}
Conversely, given a real subspace $\R$ of $\hil$, define a conjugation $\conj_\R$ by
\begin{equation}
	\label{realtoconj}	\conj_\R \sum_{i=1}^d c_i \ket[\psi_i] = \sum_{i=1}^d c_i^\ast \ket[\psi_i],
\end{equation}
where $\{ \ket[\psi_i] \}_{i=1}^d$ is a basis of $\R$. Note that $\conj_\R$ does not depend on the choice of the basis of $\R$. The mappings $\conj \mapsto \R_\conj$ and $\R \mapsto \conj_\R$ are the inverses of each other. The bijective correspondence implies that a conjugation is a complex conjugation in reference to any basis of a real subspace.

The (direct) tensor product $\antiu_A \otimes \antiu_B$ of antiunitaries $\antiu_A$ on $\hil_A$ and $\antiu_B$ on $\hil_B$ is defined by
\begin{equation}
 \label{antiunitary_tensor} \antiu_A \otimes \antiu_B \sum_j c_j \ket[\psi^A_j] \otimes \ket[\psi^B_j] := \sum_j c_j^\ast (\antiu_A \ket[\psi^A_j]) \otimes (\antiu_B \ket[\psi^B_j]),
\end{equation}
for a vector $\sum_j c_j \ket[\psi^A_j] \otimes \ket[\psi^B_j] \in \hil_A \otimes \hil_B$. $\antiu_A \otimes \antiu_B$ is again an antiunitary. This definition of the tensor product leads to $(\hat{U}_A \antiu_A) \otimes (\hat{U}_B \antiu_B) = (\hat{U}_A \otimes \hat{U}_B) (\antiu_A \otimes \antiu_B)$, where $\hat{U}_A$ and $\hat{U}_B$ are linear operators.

\subsection{Eigenvectors of antiunitary operators}
An eigenvector of an antilinear operator $\antiu$ is a vector $\ket[\eta]$ such that $\antiu \ket[\eta] = c \ket[\eta]$ holds for some complex constant $c$. Unlike for linear operators, the eigenvalues of an antilinear operator form circles in the complex plane since if $c$ is an eigenvalue of $\antiu$, $\antiu e^{i \phi} \ket[\eta] = e^{-i \phi} \antiu \ket[\eta] = e^{-2i \phi} c e^{i \phi} \ket[\eta]$.

An antilinear operator does not necessarily have an eigenvalue. Moreover, even if an antilinear operator has an eigenvalue, it does not necessarily have an eigenbasis. An antilinear operator has an eigenbasis if and only if it is Hermitian.

A representative example of antiunitaries that does not have any eigenvalue is the two-dimensional spin flip. The spin-flip operator is defined by
\begin{equation}
 \antiu_f := i \hat{\sigma}_Y \conj_\compt,
\end{equation}
where $\conj_\compt$ is complex conjugation in the computational basis and $\hat{\sigma}_Y$ is the Pauli-Y operator. A spin flip is known to operate as a universal-NOT gate, which takes any single qubit state to its orthogonal one \cite{buzekOptimalManipulationsQubits1999}. Therefore, no state is invariant under the spin-flip operation.

Eigenvectors of conjugations are related to real subspaces in the following manner:
\begin{lemma}\label{lem:real}
	The following three conditions on a vector $\ket[\eta] \in \hil$ and a conjugation $\conj$ on $\hil$ are equivalent:
	\begin{enumerate}
		\item $\ket[\eta]$ is an eigenvector of $\conj$.
		\item There exists a unimodular complex number $z$ such that $z \ket[\eta] \in \R_\conj$ holds, where $\R_\conj$ is defined by \eqref{conjtoreal}.
		\item Operator $\ketbra[\eta]$ is represented by a real matrix in any basis of $\R_\conj$. Equivalently, $\conj (\ketbra[\eta]) \conj = \ketbra[\eta]$ holds.
	\end{enumerate}
\end{lemma}
Proof.
(1 $\Leftrightarrow$ 2)
The relation $z \ket[\eta] \in \R_\conj$ in condition 2 means $\conj z \ket[\eta] = z \ket[\eta]$ by definition, and hence is equivalent to
\begin{equation}
	\label{unimodularfreedom}	\conj \ket[\eta] = z^2 \ket[\eta].
\end{equation}
This implies that $\ket[\eta]$ is an eigenvector of $\conj$ with eigenvalue $z^2$.
Conversely, if $\conj \ket[\eta] = y \ket[\eta]$ then \eqref{unimodularfreedom} holds with $z = y^{1/2}$. Eigenvalues of conjugations are unimodular complex numbers.
(2 $\Leftrightarrow$ 3)
Note that any real symmetric matrix can be diagonalized by a real orthogonal matrix through similarity transformation. Condition 3 therefore holds if and only if there exists a vector $\ket[\eta'] \in \R_\conj$ such that $\ketbra[\eta] = \ketbra[\eta']$. This is the case if and only if condition 2 is satisfied. $\qed$\\

A set of vectors $\{ \ket[f_j] \}_{j \in J}$ on a Hilbert space $\hil$ satisfying
\begin{equation}
 \label{frame} \sum_{j \in J} | \braket{x}{f_j} |^2 = \braket{x}{x} \qquad (\forall \ket[x] \in \hil),
\end{equation}
is called a \emph{frame} in harmonic analysis. In the wording of quantum information theory, a vector set $\{ \ket[f_j] \}_{j \in J}$ is a frame if and only if the set of rank-$1$ operators $\{ \ketbra[f_j] \}_{j \in J}$ is a POVM. A basis of a Hilbert space is a special kind of frame whose vectors do not overlap each other.

This article concerns the following frames related to a given conjugation.
\begin{definition}
[eigenframe] A frame, $\{ \ket[f_j] \}_{j \in J}$, is called an \emph{eigenframe} of a conjugation $\conj$ if each of its components, $\ket[f_j]$, is an eigenvector of $\conj$.
\end{definition}
A single conjugation has many different eigenframes. For example, the bases \eqref{basis2} and \eqref{basis1} are eigenframes of the same conjugation $\conj_\compt$, i.e., the complex conjugation in the two-qubit computational basis. According to lemma \ref{lem:real}, a rank-$1$ POVM $\{ \ketbra[f_j] \}_{j \in J}$ is real with respect to the subspace $\R_\conj$, if and only if $\{ \ket[f_j] \}_{j \in J}$ is an eigenframe of $\conj$.

\subsection{Matrix representation}
An indispensable tool for analyzing conjugations is their matrix representations. Let $\psi := \{ \ket[\psi_j] \}_{j=1,\ldots,d}$ be a basis of a $d$-dimensional Hilbert space $\hil$. A $d \times d$ matrix representation $[\antiu]^\psi$ of an antilinear operator $\antiu$ on $\hil$ in this basis is defined by
\begin{equation}
 \label{matrix_representation} [\antiu]^\psi_{jk} := \bra[\psi_j] \antiu \ket[\psi_k].
\end{equation}
When $[\antiu]^\psi$ is regarded as a linear operator, it satisfies
\begin{equation}
 \antiu = [\antiu]^\psi \conj_\psi,
\end{equation}
where $\conj_\psi$ is the complex conjugation in the reference basis $\psi$. The matrix $[\antiu]^\psi$ is unitary and symmetric if and only if $\antiu$ is antiunitary and Hermitian, respectively.

Two matrix representations of an antilinear operator in different bases are related by a unitary congruence transformation. Let $\eta := \{ \ket[\eta_j] \}_{j=1,...,d}$ be another basis of $\hil$, and $V$ represent the basis transformation,
\begin{equation}
 \label{interpret_takagi_vector} \ket[\psi_k] = \sum_j V_{jk} \ket[\eta_j]. \qquad (k=1,\ldots,d).
\end{equation}
Then, we obtain $[\antiu]^\eta$ from $[\antiu]^\psi$ by
\begin{equation}
 [\antiu]^\eta_{jk} = \left[ V [\antiu]^\psi V^\top \right]_{jk},
\end{equation}
where $\top$ represents transposition.

Hermitian antilinear operators have symmetric matrix representations. A symmetric matrix $A$ can be diagonalized by performing unitary congruence transformations,
\begin{equation}
 A= V ~ \mathrm{diag}(\lambda_1,\ldots,\lambda_{\dim \hil}) ~ V^\top ~~(\lambda_j \geq 0, ~\forall j)
\end{equation}
and the result is called the \emph{Autonne-Takagi factorization} of $A$. $V$'s columns and the diagonal elements $\lambda_j$ are respectively called the Takagi vectors and Takagi values of $A$. The Takagi vectors are unique up to sign if the Takagi values are non-degenerate. We call $V$, a matrix whose columns are mutually orthogonal Takagi vectors, the \emph{Takagi matrix}.

Takagi values and Takagi vectors have been used in matrix analysis for several decades. However, they are not as well known as eigenvalues and eigenvectors. Analytical techniques for computing Takagi values and vectors are detailed in \cite{bunse-gerstnerSingularValueDecompositions1988,chebotarevSingularValueDecomposition2014}, and various numerical algorithms have been proposed for their computation \cite{xuDivideconquermethod2008,xuTwistedfactorization2009,cheAdaptiveAlgorithmsComputing2018}.

A basis $\eta = \{ \ket[\eta_j] \}_{j=1,\ldots,\dim \hil}$ of $\hil$ is an eigenbasis of a Hermitian antilinear operator $\antiu$ if and only if the matrix $[\antiu]^\eta$ is diagonal. Therefore, an eigenbasis is obtained by diagonalizing $\antiu$'s matrix representation by making a unitary congruence transformation. If $V [\antiu]_\eta V^\top$ is diagonal with some basis $\eta$ and some unitary matrix $V$, an eigenbasis $\psi$ of the antilinear operator $\antiu$ can be obtained through the relation \eqref{interpret_takagi_vector}. A unitary matrix $V$ does not have to be a Takagi matrix. If $V$ is a Takagi matrix, the eigenbasis $\psi$ has the distinct property that all eigenvalues are non-negative real numbers.

\section{Local unitary equivalence classes}\label{sec:LU}
The local unitary equivalence classes of conjugations are useful for characterizing their non-local properties.
\begin{definition}
[local unitary equivalence]\label{def:lu} Two conjugations $\conj$ and $\conj'$ on the composite space $\hil_1 \otimes \cdots \otimes \hil_N$ are said to be local unitary (LU-) equivalent if there are unitary operators $\hat{U}_p$ on subsystems $\hil_p$ for all $p=1,\ldots,N$ such that
 \begin{equation}
	\label{LUsymmetry}  \left( \bigotimes_{p=1}^N \hat{U}_p \right) \conj \left( \bigotimes_{p=1}^N \hat{U}_p \right)^\dagger = \conj'.
 \end{equation}
\end{definition}
The matrix representations $[\conj]^\psi$ and $[\conj']^\psi$ of the LU-equivalent conjugations are related by a congruence transformation,
\begin{equation}
 \label{lu_repn} \left( \bigotimes_{p=1}^N U_p \right) [\conj]^\psi \left( \bigotimes_{p=1}^N U_p \right)^\top = [\conj']^\psi,
\end{equation}
where $U_p$ are now unitary matrices.

LU-equivalence is a means to characterize the non-locality of conjugation symmetries by disregarding their local properties. To exemplify this disregard for local properties, suppose we have two conjugations $\conj_1 \otimes \cdots \otimes \conj_N$ and $\conj'_1 \otimes \cdots \otimes \conj'_N$ on the same multipartite system. These two conjugations are LU-equivalent for any family of local conjugations $\conj_p$ and $\conj'_p$ ($p=1,\ldots,N$). This is analogous to the identification of all pure product states in the characterization of state entanglement.

We should also note that our version of LU-equivalence characterizes the non-locality of conjugation symmetries rather than conjugations' power of generating entanglement. Two real subspaces, $\hil_\conj$ of $\conj$ and $\hil_{\conj'}$ of $\conj'$, are interchangeable through a product unitary transformation if and only if $\conj$ and $\conj$ are LU-equivalent. In this sense, LU-equivalence pertains to the real subspaces of a tensor-product complex Hilbert space.

If our focus were on the power of conjugations to generate entanglement, our definition of LU-equivalence would have been formulated differently. Both antiunitary and unitary transformations possess the ability to convert product states into entangled states if the operators are not merely tensor products. A notable instance is the investigation conducted by Zanardi \cite{zanardiEntanglingPowerQuantum2000}, which examined the average entanglement generated by a unitary transformation applied to product states. In order to characterize the entangling power of conjugations, two conjugations should be considered equivalent if there exist product unitaries $\bigotimes_{p=1}^N \hat{U}_p$ and $\bigotimes_{p=1}^N \hat{U}'_p$ that satisfy
\begin{equation}
 \label{def:lu2} \left( \bigotimes_{p=1}^N \hat{U}_p \right) \conj \left( \bigotimes_{p=1}^N \hat{U}'_p \right) = \conj'.
\end{equation}
The corresponding symmetric unitary matrices $[\conj]^\psi$ and $[\conj']^\psi$ are related by
\begin{equation}
 \left( \bigotimes_{p=1}^N U_p \right) [\conj]^\psi \left( \bigotimes_{p=1}^N U'_p \right)^\ast = [\conj']^\psi.
\end{equation}
This particular LU-equivalence class has been solved for two-qubit unitary matrices, as demonstrated in \cite{krauscirac2001}.

This article focuses solely on the non-locality of conjugation symmetries and employs definition \ref{def:lu}, while \eqref{LUsymmetry} and \eqref{def:lu2} lead to different LU-equivalence classes.

Identifying the LU-equivalence classes of conjugations is a challenging task. Here, we concentrate on a simple system consisting of two qubits. The magic basis is a convenient two-qubit basis to work with. Section \ref{sec:canonical} introduces a canonical form of conjugations that represents each LU-equivalence class.

\subsection{LU-equivalence classes of two-qubit conjugations}
\label{sec:canonical}
LU-equivalence classes of two-qubit conjugations are characterized by combinations of four unimodular numbers defined as follows.
\begin{definition}
[magic-basis spectrum] The magic-basis spectrum of a two-qubit conjugation $\conj$ is defined to be an unordered set of $[\conj]^\magic$'s eigenvalues in which degenerate eigenvalues appear multiple times according to their degeneracy.
\end{definition}
Note the difference between the notion of a conjugations' spectrum and that of the magic-basis spectrum. A spectrum of a conjugation is an almost redundant concept since any unimodular complex number is an eigenvalue of a conjugation (see section~\ref{sec:preliminary}).

Several distinct conjugations may share the same magic-basis spectrum. This is because several distinct symmetric unitary matrices share the same spectrum. Now let us define an equivalence relation on the space of magic-basis spectra.
\begin{definition}
Two unordered sets of four unimodular complex numbers $\{ z_1,z_2,z_3,z_4 \}$ and $\{ z'_1,z'_2,z'_3,z'_4 \}$ are defined to be equivalent if there is a phase $\phi$ satisfying
 \begin{equation}
 \{ z'_1,~ z'_2,~ z'_3,~ z'_4 \} = \{ e^{i \phi} z_1,~ e^{i \phi} z_2,~ e^{i \phi} z_3,~ e^{i \phi} z_4 \},
 \end{equation}
and we denote it by $\{ z_1,z_2,z_3,z_4 \} \sim \{ z'_1,z'_2,z'_3,z'_4 \}$. The equivalence class containing $\{ z_1,z_2,z_3,z_4 \}$ is denoted by $\underline{\{ z_1,z_2,z_3,z_4 \}}$.
\end{definition}

Any two-qubit conjugation can be transformed by a suitable LU-transformation into a canonical form.
\begin{theorem}\label{thm:canonical}
Let $\conj$ be a conjugation on the two-qubit space $\hil_A \otimes \hil_B$, and let $\{ z_1, z_2, z_3, z_4 \}$ be its magic-basis spectrum. Then, there is a pair of unitaries $U$ on $\hil_A$ and $V$ on $\hil_B$ satisfying
 \begin{equation}
 \left[ \left( \hat{U} \otimes \hat{V} \right) \conj \left( \hat{U} \otimes \hat{V} \right)^\dagger \right]^\magic = \mathrm{diag} (z'_1, z'_2, z'_3, z'_4),
 \end{equation}
if and only if $\{ z'_1, z'_2, z'_3, z'_4 \} \sim \{ z_1, z_2, z_3, z_4 \}$.
\end{theorem}
Proof.
First, let us construct the unitaries $U$ and $V$.

Note that any complex symmetric unitary matrix is diagonalized by a real orthogonal matrix. If $S$ is a complex symmetric unitary matrix, then
\begin{equation}
 [\rpart [S], \ipart [S]] = \frac{[S + S^\ast , S - S^\ast]}{4i} = \frac{[S^\dagger , S] - [S, S^\dagger ]}{4i} = 0,
\end{equation}
where we have used $S^\ast = S^\dagger$ and $[S , S^\dagger] =0$. Therefore, two real Hermitian matrices $\rpart [S]$ and $\ipart [S]$ can be diagonalized by the same real orthogonal matrix. The same orthogonal matrix diagonalizes $S = \rpart [S] + i \ipart [S]$. 

Since $[\conj]^\magic$ is a complex symmetric unitary matrix, there is a real orthogonal matrix $O$ satisfying
\begin{equation}
 O [\conj]^\magic O^\top = O [\conj]^\magic O^\dagger = \mathrm{diag}(z_1,~z_2,~z_3,~z_4),
\end{equation}
where $\{ z_1,~z_2,~z_3,~z_4 \}$ is the magic-basis spectrum of $\conj$, as is introduced in the theorem.

Let $\phi$ be a phase and $\tau$ be a permutation on $\{ 1,2,3,4 \}$ such that
\begin{equation}
 (z'_1, z'_2, z'_3, z'_4) = (e^{i \phi} z_{\tau(1)},~e^{i \phi} z_{\tau(2)},~e^{i \phi} z_{\tau(3)},~e^{i \phi} z_{\tau(4)}).
\end{equation}
Let $O_\tau$ be the orthogonal matrix representing the permutation of column vectors according to $\tau$, and define a real orthogonal matrix $O'$ by
\begin{align}
 O'= \left\{ \begin{array}{lc}
 O_\tau O & (\det [O_\tau O]=1) \\
 Z_1 O_\tau O & (\det [O_\tau O]=-1)
 \end{array} \right. ,
\end{align}
where $Z_1 := \mathrm{diag}(-1,1,1,1)$. Since $\det [Z_1] = -1$, $O'$ has a positive determinant, we have
\begin{equation}
 O' [\conj]^\magic O'^\top = e^{-i \phi} \mathrm{diag}(z'_1, z'_2, z'_3, z'_4).
\end{equation}

Now let us go back and forth between two representations. There is a bijective correspondence between $4 \times 4$ orthogonal matrices with a positive determinant to a pair of single-qubit $\SU (2)$ unitary matrices given by
\begin{align}
 \label{isomorphism1} MOM^\dagger = U' \otimes V', \qquad M_{jk} := \braket{\compt_j}{\magic_k} = \frac{1}{\sqrt{2}} \left( \begin{array}{cccc}
 1 & i && \\
 && i & 1 \\
 && i & -1 \\
 1 & -i && 
 \end{array}
 \right),
\end{align}
where $O' \in \SO (4)$ and $U', V' \in \SU(2)$ \cite{hillEntanglementPairQuantum1997a,makhlin2000NonlocalPO,vatanOptimalQuantumCircuits2004}. Here, $\compt$ and $\mu$ respectively denote the two-qubit computational basis and the magic basis defined by \eqref{magic}. Let $\hat{U}$ and $\hat{V}$ be unitary operators on $\hil_A$ and $\hil_B$ which are represented by matrices $e^{i \phi/2} U'$ and $V'$ in the computational bases, respectively. When we return to the magic basis, we find
\begin{align}
 [(\hat{U} \otimes \hat{V}) \conj (\hat{U} \otimes \hat{V})^\dagger]^\magic &= [U \otimes V]^\magic [\conj]^\magic [U \otimes V]^{\magic \top} \\
 &= e^{i \phi} (M^\dagger U' \otimes V' M) [\conj]^\magic (M^\dagger U' \otimes V' M)^\top \\
 &= e^{i \phi} O' [\conj]^\magic O'^\top = \mathrm{diag}(z'_1, z'_2, z'_3, z'_4),
\end{align}
as desired.

To prove the converse, assume that
\begin{equation}
 \label{converse} \left[ \left( \hat{U} \otimes \hat{V} \right) \conj \left( \hat{U} \otimes \hat{V} \right)^\dagger \right]^\magic = \mathrm{diag} (z'_1, z'_2, z'_3, z'_4),
\end{equation}
holds for unitaries $\hat{U}$ and $\hat{V}$, where $z'_j$ ($j=1,2,3,4$) are now numbers. Let $U'$ and $V'$ be $\SU (2)$ unitary matrices and $\phi$ be a phase such that $e^{i \phi} U' \otimes V' = [\hat{U} \otimes \hat{V}]^\compt$. From \eqref{isomorphism1} and \eqref{converse} follows
\begin{equation}
 \label{converse2} O [\conj]^\magic O^\top = e^{-2i \phi} \mathrm{diag} (z'_1, z'_2, z'_3, z'_4),
\end{equation}
where $O = M^\dagger U' \otimes V' M$ is an orthogonal matrix. Since the spectrum of \eqref{converse2}'s left-hand-side $O [\conj]^\magic O^\top$ is equal to the spectrum of $[\conj]^\magic$, $\{ e^{-2i \phi} z'_1,~ e^{-2i \phi} z'_2, e^{-2i \phi} z'_3, e^{-2i \phi} z'_4 \}$ must be equal to the magic-basis spectrum. Therefore, $\{ z'_1, z'_2, z'_3, z'_4 \}$ is equivalent to the magic-basis spectrum of $\conj$.
$\qed$\\
Theorem \ref{thm:canonical} implies that any two-qubit conjugation has a magic basis $\mu'$, relative to some product basis $\compt'$, as an eigenbasis.

The complete characterization of two-qubit conjugations is obtained as an immediate consequence of theorem~\ref{thm:canonical}.
\begin{corollary}\label{cor:LU}
Two two-qubit conjugations are LU-equivalent if and only if their magic-basis spectra are equivalent.
\end{corollary}
We can label each LU-equivalence class with the representative magic-basis spectrum, i.e., $\underline{ \{ z_1, z_2, z_3, z_4 \} }$. The space of LU-equivalent classes of conjugations is homeomorphic to the quotient space of magic-basis spectra divided by $\sim$, and thus it is also homeomorphic to the configuration space of four unlabelled points on a circle.

The properties of two-qubit conjugations that are invariant under local unitary transformations can be characterized by their magic-basis spectra. One such property that we consider in section \ref{sec:measurability} is the concept of ``Prod-measurability''. In the following subsection, we prove that no property invariant under local unitary transformations can exhibit a certain bias between the two parties in the two-qubit system.

Table \ref{tab:examples} presents the known two-qubit conjugations with nonequivalent magic-basis spectra.
\begin{table}[hbtp]
\caption {Magic-basis spectra of particular two-qubit conjugations.}
\label{tab:examples}
\centering
\begin{tabular}{c|c|c}
\hline
magic-basis spectrum & conjugations & measurability \\
\hline \hline
$1,1,1,1$ & collective spin flip $\antiu_f \otimes \antiu_f$ \eqref{spin-flip} & Sep-unmeasurable \\ \hline
$1,-1,-1,-1$ & conjugate swap $\conj_\swp$ \eqref{conjugate-swap} & Sep-unmeasurable \\ \hline
$1, -1, -1, 1$ & product conjugation $\conj_A \otimes \conj_B$ \eqref{product_conjugation} & Prod-measurable \\
\hline
\end{tabular}
\end{table}
We will look more into each example in section~\ref{sec:measurability}, where we consider the ``measurability'' of conjugations.

\subsection{No bias in LU-invariant properties of two-qubit conjugations}
LU-invariant properties, in general, may exhibit a preference towards one party over another. Here, by “LU-invariant property”, we refer to a statement regarding conjugations whose veracity remains unchanged under LU transformations. For instance, there exists a set of quantum states involving two parties that can be differentiated through classical communication from one party to the other but not in the reverse direction. This one-way distinguishability, with designated senders and receivers, is an example of a biased LU-invariant property where the two parties are treated unequally.

More specifically, consider the following proposition ``$P_X(\conj)$'' ($X=A,B$) about conjugation $\conj$ on $\hil_A \otimes \hil_B$: an eigenframe measurement of $\conj$ is implementable by local operations and a classical communication from the party owning the subsystem $\hil_X$. In general, there might be a bipartite conjugation such that $P_A(\conj) = \mathrm{true}$ and $P_B(\conj) = \mathrm{false}$.

Can two-qubit conjugations exhibit a biased LU-invariant property? In other words, is there any LU-invariant property that changes its truth value for certain conjugations when the roles of the two qubit subsystems are exchanged? The following corollary provides a negative answer to this question.
\begin{theorem}
Any two-qubit conjugation $\conj$ is LU-equivalent to $\swp ~ \conj ~ \swp^\dagger$, where $\swp$ is the swap operator.
\end{theorem}
Proof.
The swap operator is represented by an orthogonal matrix $O = M^\dagger [\swp]^\compt M$ in the magic basis \cite{vatanOptimalQuantumCircuits2004}. The spectrum of $[\swp ~ \conj ~ \swp^\dagger]^\magic = O [\conj]^\magic O^\top$ is equal to the spectrum of $[\conj]^\magic$, because the orthogonal transformation does not change the spectrum. Therefore, the magic-basis spectra of $\swp \conj \swp^\dagger$ and $\conj$ are equal, and corollary~\ref{cor:LU} implies their LU-equivalence.
$\qed$\\

The LU-equivalence classes of two-qubit conjugations are preserved under permutation of the two subsystems, as are the truth values of LU-invariant properties. Therefore, if an LU-invariant property is exhibited by $\conj$, it will still be exhibited when the roles of the two qubit subsystems are exchanged.

\section{Measurability of conjugations}
\label{sec:measurability}
In this section, we investigate an LU-invariant property of conjugations which we refer to as ``measurability''. A conjugation has this property if and only if all of its LU-equivalent conjugations do so as well. As a result, measurability, as well as any other LU-invariant property, defines a partition in the space of LU-equivalence classes of conjugations.

Our focus is on the communication cost required to implement eigenframe measurements of a given multipartite conjugation. Such conjugations have eigenframes with various non-local properties, as illustrated by the two eigenframes \eqref{basis2} and \eqref{basis1} of the complex conjugation in the two-qubit computational basis. The former corresponds to a local measurement, while the latter corresponds to a Bell measurement. Measurement in eigenframe \eqref{basis2} requires fewer communication resources. The optimized communication resource depends on the properties of the multipartite conjugation.

The following LU-invariant properties will facilitate a more precise discussion on the eigenframe measurements.
\begin{definition}
A conjugation $\conj$ on a multipartite system $\hil_1 \otimes \ldots \otimes \hil_N$ is said to be Prod-measurable if the product of local frames $\left\{ \ket[f^1_{j_1}, \ldots, f^N_{j_N}] \right\}_{j_p \in J_p,~p=1,\ldots,N}$ is its eigenframe. It is said to be Sep-measurable if there is an eigenframe $\left\{ \ket[ f^1_j, \ldots f^N_j ] \right\}_{j \in J}$ comprising only product vectors.
\end{definition}

The concept of measurability is relevant to certain quantum network sensing protocols, as will be discussed in section \ref{sec:QSN}. Essential to the process of imaginarity-free quantum estimation \cite{miyazaki2022imaginarityfree}, eigenframe measurements of particular conjugations are crucial components that saturate the quantum Cram\'{e}r-Rao bound (CRB), which represents the limit of precision in parameter estimation. It is plausible that these measurements can be implemented with fewer communication resources and that one can rely on the measurability of conjugations to determine whether separable or entangled measurements are necessary.

\begin{figure}[tb]
	\centering
	\includegraphics[width=\textwidth]{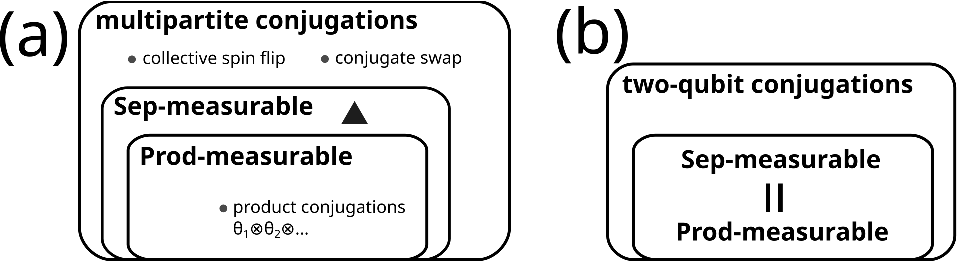}
	\caption{The measurability hierarchy of multipartite conjugations. (a) The measurability hierarchy of general multipartite conjugations with examples. The example of Prod-unmeasurable but Sep-measurable conjugation, pointed by the triangle, is given by \eqref{candidate} with basis \eqref{32basis}. (b) The measurability hierarchy of two-qubit conjugations. The classes of Sep-measurable conjugations reduces to those of Prod-measurable ones on a two-qubit system.}
	\label{fig:hierarchy}
\end{figure}

The main result of this section is the strict hierarchy of conjugations based on their measurability, which we present in figure \ref{fig:hierarchy}. We confirmed the existence of Prod-unmeasurable Sep-measurable conjugations by introducing a computable criterion for Prod-measurable conjugations. The magic-basis spectra of two-qubit conjugations directly reveal their measurability. As is shown in figure~\ref{fig:hierarchy} (b), all the Sep-measurable two-qubit conjugations are Prod-measurable.

Our primary focus will be on Prod-measurable conjugations, which will be shown to play a crucial role in phase estimation protocols of quantum metrology in section \ref{sec:QSN}. These conjugations offer an explanation for why local measurements can achieve optimal precision bounds in such protocols.

\subsection{Prod-measurable conjugations}
\label{sec:Prod-measurable}
Let us start with a primal example of Prod-measurable conjugations. Any tensor product
\begin{equation}
 \label{product_conjugation} \conj_1 \otimes \ldots \otimes \conj_N
\end{equation}
of local conjugations is Prod-measurable. Furthermore, any tensor product of local eigenframes of $\conj_p$ is an eigenframe of $\conj_1 \otimes \ldots \otimes \conj_N$. Our main subject is a quest to find more general Prod-measurable conjugations and their characterization.

The following theorem characterizes general Prod-measurable conjugations:
\begin{theorem}
[Prod-measurability]\label{thm:prod} A conjugation is Prod-measurable if and only if there is a product basis $\psi = \{ \ket[\psi_{j_1},\ldots,\psi_{j_N}] \}_{j_p=1,\ldots,\dim \hil_p}$ in which the matrix representation is diagonalized:
 \begin{equation}
 \label{product_diagonal} [\conj]^\psi_{j_1,\ldots,j_N;j_1',\ldots,j_N'} = e^{i \phi_{j_1,\ldots,j_N}} \delta_{j_1, j_1'} \ldots \delta_{j_N, j_N'},
 \end{equation}
 where $\phi_{j_1,\ldots,j_N} \in \mathbb{R}$.
\end{theorem}
The product basis $\psi = \{ \ket[\psi_{j_1},\ldots,\psi_{j_N}] \}_{j_p=1,\ldots,\dim \hil_p}$, described in the theorem, is a product eigenframe.

See appendix~\ref{appx:proof_prod} for the proof of the ``only if'' part. In the language of matrix representations, theorem \ref{thm:prod} says that a conjugation $\conj$ is Prod-measurable if $[\conj]^\compt$ is diagonalized by a product unitary matrix through a congruence transformation.

The matrix representations of product conjugations \eqref{product_conjugation} can be diagonalized by any product reference basis to the identity matrix (i.e. $e^{i \phi_{j_1, \ldots,j_N}} = 1$ for any $j_1, \ldots,j_N$). Theorem \ref{thm:prod} indicates that product conjugations are not the only Prod-measurable conjugations. For example, control-Z ($\mathrm{CZ} = {\rm diag} (1,1,1,-1)$ in the computational basis) defines a Prod-measurable conjugation.

To determine the product eigenframe, we must address the matrix diagonalization problem through the product congruence transformation. Although this seems quite challenging, the problem can be simplified by reducing it to a standard Autonne-Takagi factorization by using the subsequent corollary. Let $\trace_{\overline{p}} [ \cdot ]$ denote the partial trace of systems other than $\hil_p$ in the multipartite system $\otimes_{p'=1}^N \hil_{p'}$.
\begin{corollary}\label{cor:search}
Let $\conj$ be a Prod-measurable conjugation on $\otimes_{p=1}^N \hil_p$; then, there is a combination $(V^1, \ldots , V^N)$ of Takagi matrices $V^p$ of partial traces $\trace_{\overline{p}} \left[ [\conj]^\compt \right]$ $(p=1,\ldots ,N)$ such that $(\otimes_p V^p) [\conj]^\compt (\otimes_p V^p)^\top$ is diagonal.
\end{corollary}
Specifically, to search a product eigenframe of $\conj$, one can initially compute the Takagi matrices $U_p$ that diagonalize $\trace_{\overline{p}} \left[ [\conj]^\compt \right]$ for each $p$ through the Autonne-Takagi factorization and subsequently verify whether the product $\otimes_p V^p$ diagonalizes $[\conj]^\compt$. Nevertheless, this technique has a potential drawback in that a limitless number of Takagi matrices exist if the Takagi values are degenerate. In such instances, one may fail to identify the desired product of Takagi matrices. For example, the algorithm does not halt for the conjugate swap, which is not Prod-measurable. Appendix~\ref{appx:proof_search} gives a proof of corollary~\ref{cor:search}.

Corollary~\ref{cor:search} gives computable necessary conditions for Prod-measurable conjugations.
\begin{corollary}
[total-normality criterion]\label{cor:necessary} If a conjugation $\conj$ on $\otimes_{p=1}^N \hil_p$ is Prod-measurable, it satisfies two conditions: (i) matrices
 \begin{equation}
 \label{local_matrix} \trace_{\overline{p}} \left[ [\conj]^\compt \right],
 \end{equation}
are symmetric for all $p$, and (ii) the matrix
 \begin{equation}
 \label{matrixX} X_\conj := \left( [\conj]^\compt \right)^\dagger \bigotimes_{p=1}^N \trace_{\overline{p}} \left[ [\conj]^\compt \right],
 \end{equation}
is normal (in the sense $X_\conj X_\conj^\dagger = X_\conj^\dagger X_\conj$).
\end{corollary}
The total-normality criterion is only a necessary condition for Prod-measurability, and we know that the condition is not sufficient. Conjugation $\conj_\swp = \swp ~ \conj_\compt$, where $\swp$ is the swap operator on a $d \times d$-level system, is total-normal in the sense it satisfies conditions (i) and (ii) in corollary \ref{cor:necessary}. In section \ref{subsubsec:swap}, however, we will see that $\conj_\swp$ is not Prod-measurable. This pitfall of the total-normality criterion arises from the degenerate Takagi values of $\otimes_{p=1}^N \trace_{\overline{p}} \left[ [\conj_\swp]^\compt \right]$, for which we have the following corollary:

\begin{corollary}\label{cor:sufficient}
Let $\conj$ be a conjugation satisfying the total-normality criterion. If the matrix $\otimes_{p=1}^N \trace_{\overline{p}} \left[ [\conj]^\compt \right]$ has non-degenerate Takagi values, then $\conj$ is Prod-measurable.
\end{corollary}
Proofs of the corollaries are in appendix \ref{appx:proof_search}.

As a nontrivial example, consider a bipartite composition of a $d \times d'$ system $\mathbb{C}_d \otimes \mathbb{C}_{d'}$. Let $\conj_{d \rightarrow d'}$ be a bipartite conjugation with a block diagonal $d d' \times d d'$ unitary matrix representation,
\begin{align}
 \label{oneway} [\conj_{d \rightarrow d'}]^\compt = \left(
 \begin{array}{ccc}
 U_1 && \\
 & \ddots & \\
 && U_d
 \end{array}\right),
\end{align}
where $U_j$ are now arbitrary symmetric $d'$-level unitaries for $j=1,\ldots,d$. If $d=2$, $\conj_{d \rightarrow d'}$ satisfies the total-normality criterion for arbitrary pairs of $U_1$ and $U_2$.

If $d \geq 3$, there is a combination of $d'$-level unitaries for which $\conj_{d \rightarrow d'}$ violates the total-normality criterion. $\conj_{d \rightarrow d'}$ is not Prod-measurable for such unitaries. Section \ref{sec:sep} presents an example.

\subsubsection{Prod-measurable two-qubit conjugations}
\label{subsubsec:Prod-measurable_two-qubit}
For two-qubit conjugations, a computable exact characterization of Prod-measurability is offered by the magic-basis spectrum.
\begin{theorem}\label{thm:product_qubit}
A two-qubit conjugation $\conj$ is Prod-measurable if and only if the magic-basis spectrum is equivalent to $\{ 1,-1,z,-z \}$, where $z$ is any unimodular number. A two-qubit conjugation $\conj$ is a product conjugation if and only if the magic-basis spectrum is equivalent to $\{ 1,1,-1,-1 \}$.
\end{theorem}
A proof of this theorem is in appendix ~\ref{appx:poof_prod_meas_qubit}. The proof provides a way to find product eigenframes of Prod-measurable conjugations.

The computable characterization of Prod-measurable conjugations does not generalize to higher dimensional systems. The crucial relation $\mathrm{SU}(d) \times \mathrm{SU}(d) \simeq \mathrm{SO}(d^2)$, exhibited by the magic basis, holds only when $d=2$ \cite{bengtssonGeometryQuantumStates2017}.

\subsection{Sep-measurable conjugations}
\label{sec:sep}
Sep-measurable conjugations have eigenframes that comprise only product vectors. In this section we show that such an intermediate conjugation exists, but not in two-qubit systems.

A procedure to construct Prod-unmeasurable but Sep-measurable conjugations starts from locally indistinguishable product bases. Let $\{ \ket[\psi^A_j, \psi^B_j] \}_{j=1,\ldots,d_A \times d_B}$ be a locally indistinguishable separable basis on a $d_A \times d_B$-dimensional system. Define a conjugation by
\begin{equation}
 \label{candidate} \left( \sum_{j=1}^{d_A \times d_B} e^{i \phi_j} \ket[\psi^A_j, \psi^B_j] \bra[\psi^{A \ast}_j, \psi^{B \ast}_j] \right) \conj_\compt,
\end{equation}
where $\conj_\compt$ is the complex conjugation in the product computational basis. This conjugation is Sep-measurable for any choice of phases since the separable basis $\{ \ket[\psi^A_j, \psi^B_j] \}_{j=1,\ldots,d_A \times d_B}$ is an eigenframe. The phases $\phi_j$ ($j=1,\ldots, d_A \times d_B$) are fixed so that the conjugation becomes Prod-unmeasurable, if possible.

The procedure ends up with a Prod-unmeasurable conjugation after starting from a basis of a $3 \times 2$-dimensional system,
\begin{equation}
 \label{32basis} \ket[0_A,+_B],~\ket[0_A,-_B],~\ket[1_A,0_B],~\ket[1_A,1_B],~\ket[2_A,0_B],~\ket[2_A,1_B], 
\end{equation}
where $\ket[\pm] = (\ket[0] \pm \ket[1])/\sqrt{2}$. The conjugation is of the form $\conj_{3 \rightarrow 2}$, whose matrix representation is given by \eqref{oneway}. If we choose phases $(0,\pi,0,0,0,\pi/2)$, the resulting conjugation violates the total-normality criterion and thus is not Prod-measurable.

A construction from a locally indistinguishable basis does not always work. Here, let us start the construction from a basis of a qubit-qubit system,
\begin{equation}
 \label{22basis} \ket[0_A,+_B],~\ket[0_A,-_B],~\ket[1_A,0_B],~\ket[1_A,1_B].
\end{equation}
The four states cannot be distinguished by product measurements. However, for any choice of phases $\phi_j$ ($j=1,\ldots,4$), the conjugation \eqref{candidate} is Prod-measurable.

While the Prod-measurability of above two-qubit conjugation can be shown by an explicit calculation, we also have the following theorem:
\begin{theorem}\label{thm:no_sep}
Any two-qubit conjugations are either Prod-measurable or Sep-unmeasurable.
\end{theorem}
The proof is presented in appendix~\ref{appx:proof_product_qubit}. According to this theorem, any two-qubit conjugation constructed from a separable basis by \eqref{candidate} must be Prod-measurable. Prod-unmeasurable Sep-measurable conjugations exist only in systems with more than $2 \times 2$ dimensions.

Note that the basis \eqref{32basis} differs from \eqref{22basis} only in the two vectors $\ket[2_A, 0_B]$ and $\ket[2_A, 1_B]$. These two additional vectors do not increase the classical communication cost for discriminating the states but they do make the constructed conjugation Prod-unmeasurable. The local-distinguishability of basis elements and the Prod-measurability of conjugations are not related straightforwardly.

\subsection{Sep-unmeasurable conjugations}
Any eigenframe of a Sep-unmeasurable conjugation contains at least a single entangled eigenvector. The corresponding real subspace cannot be spanned by any set of product vectors.

We can tell if a given two-qubit conjugation is Sep-unmeasurable or not from its magic-basis spectrum (corollary~\ref{cor:LU} and theorem~\ref{thm:no_sep}). It is also possible to compute the minimum entanglement of eigenframe vectors of two-qubit conjugations. We define the average concurrence of a two-qubit frame $\mathcal{V} = \{ \ket[v_j] \}_{j=0,\ldots,n-1}$ by
\begin{equation}
		C_{ave} (\mathcal{V}) := \sum_{j=0}^{n-1} \langle v_j | v_j \rangle C(v_j), \qquad \left( C(\psi) := \frac{|\bra[\psi] \antiu_f^{\otimes 2} \ket[\psi] |}{\langle \psi | \psi \rangle} \right),
\end{equation}
where $C$ is equal to the concurrence \cite{hillEntanglementPairQuantum1997a,woottersEntanglementFormationArbitrary1998}. The concurrence of quantum states is a monotonic function of the entanglement entropy.
\begin{theorem}\label{thm:concurrence}
The minimum average concurrence for a conjugation $\conj$'s eigenframe is given by
 \begin{equation}
 \label{minimum_concurrence} \min_{\mathcal{V}:\conj's ~ \mathrm{eigenframe}} C_{ave} (\mathcal{V}) = \left| \sum_{j=1}^4 e^{i \phi_j} \right| = \left| \trace [[\conj]^\magic] \right|,
 \end{equation}
where $\{ e^{i \phi_1}, e^{i \phi_2}, e^{i \phi_3}, e^{i \phi_4} \}$ is the magic-basis spectrum of $\conj$. If $\magic$ is a canonical magic basis such that $\conj \ket[\magic_j] = e^{i \phi_j} \ket[\magic_j]$, then the vectors
 \begin{equation}
 \label{Hadamard} \ket[v_k] := \sum_j \frac{H_{jk}}{2} e^{i \phi_j /2} \ket[\magic_j], \qquad k=1,\ldots,4,
 \end{equation}
share the same concurrence $C(v_k) = |\sum_{j=1}^4 e^{i \phi_j} |/4$ and form an eigenframe attaining the minimum average concurrence. Here, $H$ is a $4 \times 4$ real Hadamard matrix.
\end{theorem}
The frame \eqref{Hadamard} minimizes the entanglement entropy on average as well as the average concurrence, since the former is a convex function of the latter. (See appendix \ref{appx:entanglement_two-qubit} for more details on this point and the proof of theorem~\ref{thm:concurrence}.)

\begin{figure}[tb]
\centering
\includegraphics[width=8.3cm]{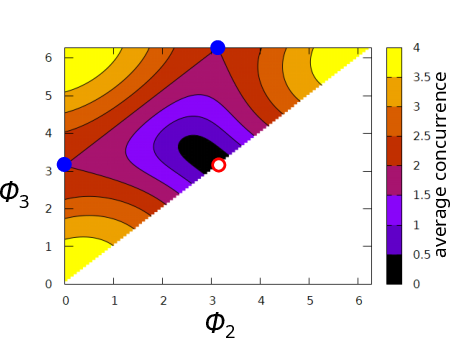}
\caption{Minimum average concurrence of eigenframes of two-qubit conjugations with magic-basis spectra $\{ 1,1, e^{i \phi_2}, e^{i \phi_3} \}$ depicted in the area $0 \leq \phi_2 \leq \phi_3 \leq 2 \pi$. The average concurrence has non-zero values except at $\phi_2 = \phi_3 = \pi$ (pointed by the red circle), where the corresponding conjugation is Prod-measurable. The magic-basis spectra at the two points of blue dots are equivalent to $\{ 1,-1,-1,-1 \}$.}
\label{fig:concurrence2}
\end{figure}

Figure \ref{fig:concurrence2} depicts the minimum average concurrence \eqref{minimum_concurrence} of conjugations.
It is worth noting that several LU-nonequivalent conjugations can have the same minimum average concurrence of eigenframes. For instance, conjugations with magic-basis spectra $\{ 1,1,1,e^{i \pi} \}$ and $\{ 1,1, e^{i 2 \pi/3}, e^{i 2 \pi/3} \}$, despite not being LU-equivalent, have the same minimum average concurrence of $2$.

The value of $\min C_{ave}/4$, obtained by dividing the minimum average concurrence by $4$, serves as a lower bound of the states' concurrence required to deterministically implement a symmetric measurement. To elaborate, the minimum average concurrence is attained by the four vectors sharing the same concurrence. To generate any of the four states as a measurement outcome, an entangled state with a concurrence of at least $\min C_{ave}/4$ is required. In the case that the vectors of a different eigenframe do not share the same concurrence, the concurrence of one of the components exceeds $\min C_{ave}/4$, necessitating a greater degree of entanglement for its production. Thus, it becomes necessary to have an ancillary state supply with a concurrence of at least $\min C_{ave}/4$.

Note that $\min C_{ave}/4$ may not be sufficient to implement an eigenframe measurement. Determining the minimum entanglement for a measurement implementation is a difficult task, even for two-qubit systems \cite{bandyopadhyayEntanglementCostNonlocal2009,bandyopadhyay2010}. The authors looked for a method to implement the elegant joint measurement \cite{gisinEntanglement25Years2019,czartowskiBipartiteQuantumMeasurements2021} with an ancillary input state with concurrence $1/2$, but were unable to find one. The vectors of an elegant joint measurement form an eigenframe of $\conj_\swp$ (defined in section \ref{subsubsec:swap}) saturating $\min C_{ave}/4 = 1/2$. On the basis of our efforts so far, we conjecture that $\min C_{ave}/4$ serves as a lower bound that is not saturated.

Below, we look into more detail at two representative examples of two-qubit conjugations whose magic basis spectra are $(1,1,1,1)$ and $(1,-1,-1,-1)$. Then, we consider generalizations of the two Sep-unmeasurable qubit-qubit conjugations to higher dimensional systems. The generalized conjugations remain Sep-unmeasurable and inherit the distinct properties of the original qubit-qubit conjugations.

\subsubsection{(1,1,1,1): collective spin flips}
\label{subsubsec:spin-flip}
Our first example of a Sep-unmeasurable conjugation has the magic-basis spectrum $(1,1,1,1)$. It is the tensor product $\antiu_f^{\otimes 2}$ of spin flips on a two-qubit space. Its matrix representation is given by
\begin{align}
 \label{spin-flip} [\antiu_f^{\otimes 2}]^\compt = [i \sigma_y \otimes i \sigma_y]^\compt
 = \left(
 \begin{array}{cccc}
 & & 0 & 1\\
 & & -1 & 0\\
 0 & -1 & & \\
 1 & 0 & &
 \end{array}\right),
\end{align}
in the product computational basis $\compt$.

A two-qubit state is an eigenvector of $\antiu_f^{\otimes 2}$ if and only if it is maximally entangled. The Bell measurement is an eigenframe measurement.

The magic basis \eqref{magic} is a reference basis of $\antiu_f^{\otimes 2}$. Any real linear combination of vectors from the magic basis is thus an eigenvector of $\antiu_f^{\otimes 2}$, and is maximally entangled \cite{bennettMixedStateEntanglement1996}.

It is remarkable that the collective spin flip is a conjugation while the single spin flip itself is not. The collective spin flip and its higher-dimensional generalizations are the only examples of this kind. If a tensor product of two antiunitary operators is a conjugation, they are either both conjugations or unitarily equivalent to direct sums of spin-flip operators (see appendix~\ref{sec:conjugationfromantiunitary} for details). If $\antiu_1$ and $\antiu_2$ are unitarily equivalent to direct sums of spin flips, then $\antiu_1 \otimes \antiu_2$ is Sep-unmeasurable, since neither $\antiu_1$ nor $\antiu_2$ has any eigenvector.

One may also wonder if a tensor product of three or more non-hermitian antiunitary operators can be a conjugation. It turns out that there is no such ``genuine multipartite conjugation'' that can be made from local antiunitaries (see appendix~\ref{sec:conjugationfromantiunitary} for details).

\subsubsection{(1,-1,-1,-1): conjugate swap}
\label{subsubsec:swap}
Let $\swp$ be the swap operator on the tensor product of two $d$-dimensional spaces $\hil \otimes \hil$. We define a conjugation $\conj_\swp$ by
\begin{equation}
 \label{conjugate-swap} \conj_\swp := \swp ~ \conj_\compt,
\end{equation}
and call it the conjugate swap.

When $d=2$, the two-qubit conjugate swap has the magic-basis spectrum $\{ 1, -1, -1, -1 \}$, implying Sep-unmeasurability. In figure \ref{fig:concurrence2}, the blue dots indicate the magic-basis spectra equivalent to $\{ 1, -1, -1, -1 \}$. Unlike the collective spin flip, the two-qubit conjugate swap has product eigenvectors, such as $\ket[0,0]$ and $\ket[1,1]$. Any eigenframe of the conjugate swap, however, must contain at least a single entangled vector. Among the eigenframes of the conjugate swap, ones consisting only of vectors proportional to
\begin{align}
	\label{ejv_2} \ket[\Psi_\vecn] &:= \frac{1+ \sqrt{3}}{2\sqrt{2}} \ket[\vecn,\vecn^\ast] + \frac{1- \sqrt{3}}{2\sqrt{2}} i \sigma_Y \otimes i \sigma_Y \ket[\vecn^\ast , \vecn] \\
 &= \frac{1}{2} \ket[\Psi_+] + \frac{\sqrt{3}}{2} \sin \theta \cos \phi \ket[\Psi_-] + \frac{\sqrt{3}}{2} \cos \theta \ket[\Phi_+] - \frac{\sqrt{3}}{2} i \sin \theta \sin \phi \ket[\Phi_-], \\
	& \left( \ket[\vecn] = \cos \frac{\theta}{2} \ket[0] + e^{i \phi} \sin \frac{\theta}{2} \ket[1] \right),
\end{align}
exhibit the minimum average concurrence of eigenframe vectors (see appendix~\ref{appx:ejm-minimum} for the proof). In \eqref{ejv_2}, $\vecn = ( \sin \theta \cos \phi ,~ \sin \theta \sin \phi,~ \cos \theta )$ is a directional vector pointing on the unit sphere. The vector $\ket[\Psi_\vecn]$ is an eigenvector of $\conj_\swp$ with eigenvalue $1$ for any $\vecn$. In the special case where $\vecn$ points to the four vertices of a tetrahedron inscribed in the unit sphere, the corresponding eigenframe measurement is local unitary equivalent to the elegant joint measurement \cite{gisinSpinFlipsQuantum1999,gisinEntanglement25Years2019}.

The conjugate swap is Sep-unmeasurable for any dimension $d$. Otherwise, the precision limit of multiparameter estimations from parallel states $\hat{\rho}_{\bm x} \otimes \hat{\rho}_{\bm x}$ and from mutually conjugate states $\hat{\rho}_{\bm x} \otimes \hat{\rho}_{\bm x}^\ast$ would have to coincide for any $\hat{\rho}_{\bm x}$, but this is not always the case \cite{miyazaki2022imaginarityfree}. A higher-dimensional generalization of the elegant joint measurement is constructed in \cite{czartowskiBipartiteQuantumMeasurements2021} under the assumption that SCI-POVMs exist for each dimension. The generalized elegant joint measurement defines a reference basis of $\conj_\swp$.

\section{Conjugations in quantum sensor networks}
\label{sec:QSN}
Here, we examine the conjugation symmetries of quantum sensor networks (QSNs) for imaginarity-free estimations. QSNs aim to estimate physical parameters accurately by using quantum systems shared by multiple nodes. A typical QSN protocol involves three steps: preparing the initial state, evolving the state to be estimated, and measuring the state. Theoretically, entanglement in the initial states augments the precision for specific QSNs \cite{proctorMultiparameterEstimationNetworked2018}. Our interest is in the entanglement required for the measurement step.

In several situations, entangled measurements do not enhance precision. Local operations and classical communication (LOCC) measurements saturate the optimal precision limit for single-parameter estimations with pure states or certain rank-two mixed states \cite{zhouSaturatingQuantumCramer2020}. Although phase estimation requires an entangled initial state, local measurements are sufficient to achieve the optimal precision \cite{giovannettiQuantumMetrology2006,giovannettiAdvancesQuantumMetrology2011}. In this section, we exemplify several kinds of imaginarity-free QSNs, including the phase estimation.

Imaginarity-free estimation, as introduced in \cite{miyazaki2022imaginarityfree}, employs states and measurements that exhibit symmetry under conjugations. The measurements achieving the optimal precision limit are straightforwardly derived from the conjugation and possess a degree of flexibility that could prove advantageous for QSNs. The non-local resources necessary for measurements in QSNs are dictated by conjugations, or more precisely, by the measurability of conjugations discussed in section \ref{sec:measurability}.

The discussion begins in section \ref{subsec:imaginarity-free} below with a review of imaginarity-free estimation.

One of the authors previously used antiunitary symmetry to demonstrate that bipartite entangled measurements saturate the precision bound in 3D-magnetometry \cite{baumgratzdatta2016} of an arbitrary size \cite{miyazaki2022imaginarityfree}. In section \ref{subsec:magnetometry}, we utilize a more systematic methodology on the same model to identify the local conjugation symmetries and their corresponding measurements. Our procedure replicates the established optimal measurements for phase estimation and 3D magnetometry as particular cases.

In section \ref{subsec:antiparallel}, we propose an idea for manipulating the distribution of entangled measurements in a general QSN. This idea involves doubled local systems at all nodes and the ``antiparallel model'' proposed in \cite{miyazaki2022imaginarityfree}. Here, measurements that are entangled between nodes of the original QSN are replaced by those entangled inside the nodes of the antiparallel model.

\subsection{Saturating the quantum Cram\'{e}r-Rao bound by using eigenframe measurements}\label{subsec:imaginarity-free}
Let $\{ \hat{\rho}_{\bm x} | {\bm x} \in X \subset \real^n \}$ be a quantum statistical model, that is, a set of parametrized quantum states. While we abstain from presupposing specific processes for generating the parametrized states at this juncture, one can envision a scenario wherein a state $\hat{\rho}_{\bm x} = \hat{U}_{\bm x} \hat{\rho}_{ini} \hat{U}_{\bm x}^\dagger$ is derived from an initial state $\hat{\rho}_{ini}$ following a parametrized unitary evolution $\hat{U}_{\bm x}$ by way of illustration. We will focus our attention on particular models that have conjugation symmetries:
\begin{definition}
	A quantum statistical model $\{ \hat{\rho}_{\bm x} | {\bm x} \in X \subset \real^n \}$ is said to be imaginarity-free if there is a conjugation $\conj$ such that
	\begin{equation}
		\conj \hat{\rho}_{\bm x} \conj = \hat{\rho}_{\bm x},
	\end{equation}
	holds for any ${\bm x} \in X$.
\end{definition}
The aim of section \ref{sec:QSN} is to exemplify imaginarity-free models on QSNs together with their eigenframe measurements. The pair has a distinct property in regard to parameter estimation, as outlined below (see \cite{miyazaki2022imaginarityfree} for details).

The goal of quantum parameter estimation is to achieve the ultimate precision limit in guessing the parameter value ${\bm x}$ from measurement outcomes on parameterized quantum states $\hat{\rho}_{\bm x}$. Several optimality criteria of measurements should be employed depending on the case: here, we focus on measurements that maximize the classical Fisher information, and consider the case of a POVM measurement $\Pi = \{ \hat{\Pi}_\omega \}_{\omega \in \Omega}$ on the single copy of a quantum state $\hat{\rho}_{\bm x}$. We define the $n \times n$ classical Fisher information matrix $\mathcal{F}_C (\Pi, {\bm x})$ by
\begin{equation}
	[\mathcal{F}_C (\Pi, {\bm x})]_{ij} := \sum_{\omega \in \Omega} \frac{\partial_i \trace[\hat{\Pi}_\omega \hat{\rho}_{\bm x}] \partial_j \trace[\hat{\Pi}_\omega \hat{\rho}_{\bm x}]}{\trace[\hat{\Pi}_\omega \hat{\rho}_{\bm x}]},
\end{equation}
where $\partial_i$ denotes the partial derivative with respect to $x_i$. The classical Fisher information matrix satisfies the quantum Cram\'{e}r-Rao bound (QCRB)
\begin{equation}
	\label{QCRB} \mathcal{F}_C (\Pi, {\bm x}) \leq \mathcal{F}_Q ({\bm x}),
\end{equation}
where $\mathcal{F}_Q ({\bm x})$ is the quantum Fisher information matrix
\begin{equation}
	[\mathcal{F}_Q ({\bm x})]_{ij} := \text{Re} [\trace[\hat{\rho}_{\bm x} \hat{L}_{{\bm x}, i} \hat{L}_{{\bm x}, j}]],
\end{equation}
at point ${\bm x}$ \cite{helstromMinimumVarianceEstimates1968,helstrom1976,holevo1982}. Hermitian linear operators $\hat{L}_{{\bm x}, i}$ are symmetric logarithmic derivertives defined indirectly by
\begin{equation}
	2 \partial_i \hat{\rho}_{\bm x} = \hat{L}_{{\bm x}, i} \hat{\rho}_{\bm x} + \hat{\rho}_{\bm x} \hat{L}_{{\bm x}, i},
\end{equation}
for $i=1,\ldots,n$, and are not necessarily unique.
\begin{definition}\label{QCRB-saturating}
	A POVM is said to saturate the QCRB of model $\{ \hat{\rho}_{\bm x} | {\bm x} \in X \subset \real^n \}$ at ${\bm x}$ if \eqref{QCRB} holds with equality at ${\bm x}$, namely, if the classical Fisher information of the POVM is equal to the quantum Fisher information.
\end{definition}
QCRB-saturating measurements exist for single parameter models ($n=1$), but do not necessarily exist for multiparameter models ($n \geq 2$). See \cite{szczykulskaMultiparameterQuantumMetrology2016} for a review on multiparameter precision bounds.

An approach to discover QCRB-saturating measurements, if they exist, involves computing the basis that concurrently diagonalizes symmetric logarithmic derivatives for all parameters. However, the symmetric logarithmic derivatives themselves become more challenging to compute as the system size increases. Particularly, one must seek combinations of symmetric logarithmic derivatives that mutually commute for this purpose. No systematic method is known to search for such combinations.

\begin{figure}[tb]
	\centering
	\includegraphics[width=.25\textwidth]{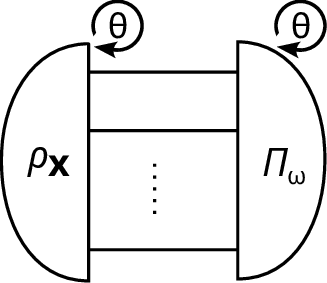}
	\caption{A circuit diagram illustrating imaginarity-free estimation on QSNs. Each line denotes a subsystem of a quantum sensor within the network. The general imaginarity-free estimation employs the model $\{ \hat{\rho}_{\bm x} | {\bm x} \in X \}$ and POVM $\{ \hat{\Pi}_\omega \}_{\omega \in \Omega}$ which are symmetric under the same conjugation $\conj$.}
	\label{fig:ifree}
\end{figure}

The following theorem demonstrates the existence of QCRB-saturating measurements for imaginary-free pure state models, along with their straightforward construction.
\begin{theorem}\label{thm:QCRB-saturating}
	Let $\{ \hat{\rho}_{\bm x} | {\bm x} \in X \subset \real^n \}$ be a pure state model such that there exists a conjugation $\conj$ satisfying $\conj \hat{\rho}_{\bm x} \conj = \hat{\rho}_{\bm x}$ for any ${\bm x} \in X$. A POVM measurement in any eigenframe of $\conj$ saturates the QCRB everywhere in $X$.
\end{theorem}
We will omit the proof of this theorem since it directly follows from theorem 3 of \cite{miyazaki2022imaginarityfree} and the definition of eigenframes. In short, any eigenframe measurement of a pure imaginarity-free model saturates its QCRB at any parameter value. ``\emph{Imaginarity-free estimation}'' from \cite{miyazaki2022imaginarityfree} refers to quantum estimation techniques utilizing imaginarity-free models and the measurements symmetric under the same conjugation, as illustrated in figure~\ref{fig:ifree}.

We adhere to imaginary-free estimation due to the versatility of QCRB-saturating measurements. As Theorem \ref{thm:QCRB-saturating} asserts, any frame from the conjugation-symmetric real subspace saturates the QCRB. This flexibility enables the selection of measurements based on available resources.

When estimating the parameters encoded in a QSN it is often desirable to use measurement that can be implemented using a reduced number of few non-local resources. Zhou et al.\cite{zhouSaturatingQuantumCramer2020} partly solved this problem by constructing QCRB-saturating LOCC measurements for all single-parameter estimations from multipartite pure states. In the multiparameter scenario, the measurements must be entangled in certain QSNs such as those consisting of antiparallel spins \cite{gisinSpinFlipsQuantum1999,changetal2014,miyazaki2022imaginarityfree}. Motivated by theorem \ref{thm:QCRB-saturating}, we will explore QCRB-saturating measurements of several conjugation-symmetric QSN models by applying our analysis on measurability (section \ref{sec:measurability}).

Before proceeding, we should note that we are interested only in the measurement saturating the QCRB, and do not care if the measurement enables full parameter estimation. Thus it may be impossible to estimate the parameter value from the outcomes of deduced measurement. To illustrate, the binary measurement in basis $(\ket[0] \pm \ket[1])/\sqrt{2}$ saturates the QCRB for any phase $\phi \in [-\pi, \pi]$ in the model $(\ket[0] \pm e^{-i \phi} \ket[1])/\sqrt{2}$. However, this measurement cannot distinguish the state with phase $+\phi$ and $-\phi$, and thus is impotent for estimating $\phi$ around $0$. We will not delve into this problem here as our primary goal is finding any QCRB-saturating measurement. Conjugation-symmetric models can overcome this problem through the use of special measurements (corollary 1 of \cite{miyazaki2022imaginarityfree}), and this idea should be adapted to QSNs in future work.

\subsection{Example: magnetometry}\label{subsec:magnetometry}
We consider conjugation symmetries of $N$-qubit networks for sensing multidimensional fields, following \cite{baumgratzdatta2016}. Let us identify each qubit with a spin-$1/2$ system that is exposed to a uniform magnetic field. We will consider three types of magnetic field with different spatial dimensions: a magnetic field along the $Z$-axis, one in the $XZ$-plane, and one that may point at any direction in three-dimensional space. The system is prepared in an initial pure state $\hat{\rho}_{ini} = \ketbra[\psi_{ini}]$ and undergoes unitary evolution
\begin{align}
	\hat{\rho}_{ini} \mapsto \hat{U} \hat{\rho}_{ini} \hat{U}^\dagger, \qquad \hat{U} = \left\{ \begin{array}{ll}
		e^{i \phi_Z \hat{H}_Z} & (\mathrm{1-dimension}) \\
		e^{i ( \phi_Z \hat{H}_Z + \phi_X \hat{H}_X)} & (\mathrm{2-dimension}) \\
		e^{i ( \phi_Z \hat{H}_Z + \phi_Y \hat{H}_Y + \phi_Y \hat{H}_Y)} & (\mathrm{3-dimension})
	\end{array}\right.,
\end{align}
where
\begin{align}
	\hat{H}_K := \sum_{p=1}^N \hat{\id}_1 \otimes \cdots \otimes \hat{\id}_{p-1} \otimes \hat{\sigma}_K \otimes \hat{\id}_{p+1} \otimes \cdots \otimes \hat{\id}_N \qquad (K=X,Y,Z),
\end{align}
and where the parameters $\phi_X, \phi_Y,$ and $\phi_Z \in [-\pi/2,\pi/2)$ indicate the strength of the magnetic field in the respective directions. The state of the model is given by $\hat{U} \hat{\rho}_{ini} \hat{U}^\dagger$. The model may or may not be conjugation-symmetric depending on the initial state.

\begin{figure}[tb]
	\centering
	\includegraphics[width=.6\textwidth]{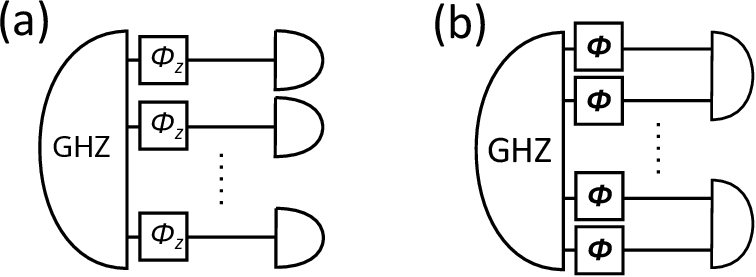}
	\caption{Circuit diagrams for (multi)dimensional field estimation. An $n$-qubit sensor network is coupled with a uniform field whose intensity is to be determined.
    (a) When the field aligns with the $Z$-axis, commencing from the initial GHZ state \eqref{GHZK} ($K=Z$), each qubit undergoes the unitary evolution $e^{- i \phi_z \hat{\sigma}_z}$. The system culminates in a final state parameterized by $\phi_z$, symmetric under an $n$-tensor product of single-qubit conjugation $\conj$ as described in \eqref{solution1}. The QCRB for parameter $\phi_z$ is hence achieved through specific product measurements.
    (b) In cases where the spatial dimension of the field is two or three, each qubit instead experiences the unitary evolution $e^{- i \sum_K \phi_K \hat{\sigma}_K}$, where the sum pertains to the relevant spatial axis. If the system is initialized to the superposed GHZ state \eqref{superposedGHZ}, the parameterized final state exhibits symmetry under an $n/2$-tensor product of bilocal conjugation $\conj$, as defined in \eqref{solution2}. The QCRB for the multidimensional field is thereby achieved via certain bilocal measurements.}
	\label{fig:magnetometry}
\end{figure}

To find conjugation-symmetries with less non-locality, we suppose two levels of locality ansatzes on the conjugation, and check if a solution exists for each dimension of the magnetic field. Here, let us denote by $\{ A, B \}$ the anticommutator $AB+BA$ of linear or antilinear operators $A$ and $B$.
\begin{description}
	\item[Ansatz1] There exists a qubit conjugation $\conj$ satisfying
	\begin{description}
		\item[(G)] $\{ \conj, \hat{\sigma}_K \} = 0$, where $K=Z$ for a one-dimensional field, $K=Z,X$ for a two-dimensional field and $K=X,Y,Z$ for a three-dimensional field, and
		\item[(S)] $\ket[\psi_{ini}]$ is an eigenvector of $\conj^{\otimes N}$.
	\end{description}
\end{description}
\begin{description}
	\item[Ansatz2] There exists a two-qubit conjugation $\conj$ satisfying
	\begin{description}
		\item[(G)] $\{ \conj, \hat{\sigma}_K \otimes \hat{\id} + \hat{\id} \otimes \hat{\sigma}_K \} = 0$, where $K=Z$ for a one-dimensional field, $K=Z,X$ for a two-dimensional field and $K=X,Y,Z$ for a three-dimensional field, and
		\item[(S)] $\ket[\psi_{ini}]$ is an eigenvector of $\conj^{\otimes N/2}$.
	\end{description}
\end{description}
Ansatz2 is considered only when $N$ is an even number for the conjugation $\conj^{\otimes N/2}$ in ``S'' to be well-defined. Conditions ``S'' are imposed solely on the initial state, whereas ``G'' are only on the generators of the state evolution. It is possible to construct local and bilocal conjugation-symmetries of the model from the solutions of Ansatz1 and Ansatz2, respectively:
\begin{lemma}
	(1) If Ansatz1 has a solution, then so does Ansatz2. (2) If qubit conjugation $\conj$ is a solution of Ansatz1, then the model with state $\hat{U} \hat{\rho}_{ini} \hat{U}^\dagger$ is symmetric under $\conj^{\otimes N}$. (3) If two-qubit conjugation $\conj$ is a solution of Ansatz2, then the model with state $\hat{U} \hat{\rho}_{ini} \hat{U}^\dagger$ is symmetric under $\conj^{\otimes N/2}$.
\end{lemma}
Proof.
(1) Let a single-qubit conjugation $\conj$ be a solution of Ansatz1. Then $\conj^{\otimes 2}$ is a solution of Ansatz2. (2)(3) Let $\conj'$ be either $\conj^{\otimes N}$ for the solution of Ansatz1 or $\conj^{\otimes N/2}$ for the solution of Ansatz2. Ansatz1-G and Ansatz2-G each imply that $\{ \conj', \hat{H}_K \} = 0$, where $K$ only takes the relevant values. This implies
\begin{equation}
	\label{invariant-generator}	\conj' \hat{U} \conj' = e^{\conj' i \left( \sum_K \phi_K \hat{H}_K \right) \conj'} = e^{- i \left( \sum_K \phi_K \conj' \hat{H}_K \conj' \right)} = e^{i \left( \sum_K \phi_K \hat{H}_K \right)} = \hat{U}.
\end{equation}
Ansatz1-S and Ansatz2-S each imply $\conj' \ket[\psi_{ini}] \propto \ket[\psi_{ini}]$, which is equivalent to
\begin{equation}
	\label{invariant-state}	\conj' \hat{\rho}_{ini} \conj' = \hat{\rho}_{ini}.
\end{equation}
Equations \eqref{invariant-generator} and \eqref{invariant-state} together imply
\begin{equation}
	\conj' \hat{U} \hat{\rho}_{ini} \hat{U}^\dagger \conj' = \conj' \hat{U} \conj' \conj' \hat{\rho}_{ini} \conj' (\conj' \hat{U} \conj')^\dagger = \hat{U} \hat{\rho}_{ini} \hat{U}^\dagger,
\end{equation}
namely, that the model is symmetric under $\conj'$.
$\qed$\\
Statement (1) of the lemma says that Ansatz1 is stricter than the second one in this sense. We indeed see that, for two- and three-dimensional magnetic fields, only the second ansatz has a solution. Below, we summarize the solutions obtained by simple matrix calculations and techniques from section \ref{sec:measurability}, in the form of a theorem.
\begin{theorem}
	\begin{enumerate}
		\item For one-dimensional field ($K=Z$), a qubit conjugation $\conj$ satisfies Ansatz1-G if and only if its matrix representation is given by \eqref{solution1}. There is no single qubit conjugation satisfying Ansatz1-G for two- and three-dimensional fields.
		\item A two-qubit conjugation $\conj$ satisfies Ansatz2-G for a two-dimensional field ($K=X,Z$) if and only if its matrix representation is given by \eqref{solution2}. The two-qubit conjugation is not Sep-measurable and must have the mininum averaged concurrence no-less than $2$.
		\item A two-qubit conjugation $\conj$ satisfies Ansatz2-G for a three-dimensional field ($K=X,Y,Z$) if it does so for a two-dimensional field.
	\end{enumerate}
\end{theorem}
The solutions of the ansatz yield the imaginarity-free estimation of multidimensional fields, with their circuit diagrams depicted in \ref{fig:magnetometry}.

Example solutions of the ansatz are presented in Table \ref{tab:magnetometry} and are detailed in the following subsection, where we set the eigenvectors of the Pauli matrices to
\begin{equation*}
	\ket[Z_+] := \ket[\compt_0], ~ \ket[Z_-] := \ket[\compt_1], ~
	\ket[X_\pm] := \frac{\ket[\compt_0] \pm \ket[\compt_1]}{\sqrt{2}}, ~ \ket[Y_+] := \frac{\ket[\compt_0] + i \ket[\compt_1]}{\sqrt{2}}, ~ \ket[Y_-] := \frac{i \ket[\compt_0] + \ket[\compt_1]}{\sqrt{2}},
\end{equation*}
and define the GHZ state-vector by
\begin{equation}
	\label{GHZK}	\ket[\mathrm{GHZ}_K] := \frac{1}{\sqrt{2}} \left( \ket[K_+]^{\otimes N} + \ket[K_-]^{\otimes N} \right),
\end{equation}
where $K=X,Y,Z$.
\begin{table}[t]
	\centering
	\begin{tabular}{|c|c|c|c|c|c|}\cline{3-6}
	\multicolumn{2}{c|}{} & \multicolumn{2}{c|}{Ansatz1} & \multicolumn{2}{c|}{Ansatz2} \\
	\cline{3-6}
	\multicolumn{2}{c|}{} & conjugation & initial state & conjugation & initial state \\ \hline
	dimen-			& $1(Z)$ & $\hat{\sigma}_X \conj_\compt$ & $\ket[GHZ_Z]$ & & \\ \cline{2-6}
	sion	& $2(X,Z),$	& No	&	& $\antiu_f^{\otimes 2}$, & $\ket[GHZ_K]$, \\
				& $3(X,Y,Z)$		&		&	& $\swp ~ \antiu_f^{\otimes 2}$ & superposed GHZ \eqref{superposedGHZ} \\ \hline
	\end{tabular}
	\caption{Conjugation-symmetries and initial states for magnetometry obtained by solving ansatz 1 and 2.}
	\label{tab:magnetometry}
\end{table}

\subsubsection*{Case of one-dimensional field}
Ansatz1 has solutions for one-dimensional field. The qubit conjugation $\conj$ has the following matrix representation:
\begin{align}
	\label{solution1}	[\conj]^\compt = c~ \sigma_X,
\end{align}
where $c$ can be any unimodular complex number.	
We can ignore this constant $c$ when the conjugation acts only in the form $\conj (\cdot) \conj$, since $c \conj (\cdot) c \conj = c c^\ast \conj (\cdot) \conj = \conj (\cdot) \conj$.

An important initial state satisfying Ansatz1-S for the above qubit conjugation is the GHZ state given by \eqref{GHZK} (K=Z). The imaginarity-free estimation of one-dimensional field, relying on the GHZ initial state and the eigenframe measurement of $\conj^{\otimes N}$, is illustrated in circuit diagram (a) of figure \ref{fig:ifree}. When the initial state is the GHZ state, the resulting model is for the celebrated phase estimation protocol, in which a product measurement is known to saturate the QCRB \cite{giovannettiQuantumMetrology2006,giovannettiAdvancesQuantumMetrology2011}. This accords with our expectation from the symmetry under product conjugation.

\subsubsection*{Cases of two- and three-dimensional fields}
The first ansatz does not have a solution for two-dimensional field: the qubit conjugation \eqref{solution1} cannot further satisfy $\conj \hat{\sigma}_X \conj = - \hat{\sigma}_X$.

Here we will take $N$ to be even and provide a general solution of Ansatz2-G for a two-dimensional field, which coincides with the solution for a three-dimensional field. The two-qubit conjugation $\conj$ has the matrix representation
\begin{align}
	\label{solution2}	[\conj]^\compt = c [\conj (\alpha)]^\compt = c \left( \begin{array}{cccc}
								&&& 1 \\
								& - e^{i \alpha} \cos \alpha & i e^{i \alpha} \sin \alpha & \\
								& i e^{i \alpha} \sin \alpha & - e^{i \alpha} \cos \alpha & \\
								1 &&&
							\end{array}\right),
\end{align}
with a unimodular complex number $c$ and a real number $\alpha$.

The methodologies elucidated in preceding sections \ref{sec:LU} and \ref{sec:measurability} prove to be useful for scrutinizing the measurability of conjugation \eqref{solution2}. We find the representation of $\conj$ in the magic basis $\mu$
\begin{equation}
	\label{solution2-magic}	[\conj]^\mu = c [\conj (\alpha)]^\mu = c M^\dagger [\conj (\alpha)]^\compt M^\ast = c ~ \text{diag}(1,1,1,-e^{2 i \alpha}),
\end{equation}
which has already been diagonalized (matrix $M$ is delineated in \eqref{isomorphism1}). The magic-basis spectrum is equivalent to $\{ 1,1,1,-e^{2 i \alpha} \}$. Given that the magic-basis spectrum does not align with $\{ 1,-1,z,-z \}$ for any unimodular $z$ and any $\alpha$, theorems \ref{thm:product_qubit} and \ref{thm:no_sep} deduce that the two-qubit conjugation $\conj$ is not Sep-measurable for any $\alpha$.

Let us deduce eigenframe measurements of $\conj$. The $N/2$-tensor products of the eigenframe measurement attain the QCRB, provided that the conjugation additionally satisfies Ansatz2 (S). The already diagonalized representation \eqref{solution2-magic} implies that the magic basis $\magic$ serves as an eigenbasis of $\conj$ for any value of $\alpha$. To investigate other eigenframes, it is noteworthy that the minimum averaged concurrence for $\conj$'s eigenframes is given by
\begin{equation}
	\label{concurrence-alpha}	\min_{\mathcal{V}:\conj's ~ \mathrm{eigenframe}} C_{ave} (\mathcal{V}) = 2 \sqrt{1 + 3 \sin^2 \alpha},
\end{equation}
as per theorem \ref{thm:concurrence} applied to the magic-basis spectrum $\{ 1,1,1,-e^{2 i \alpha} \}$. Given that the minimum averaged concurrence is below $4$ when $\alpha \neq \pm \pi /2$, there exist less-entangled eigenframes than the magic basis in this domain.
Notably, for each $\alpha$ the iso-entangled basis introduced in \cite{tavakoliBilocalBellInequalities2021}
\begin{equation}
	\label{generalized-ejv} \frac{\sqrt{3} + e^{i \alpha}}{2\sqrt{2}} \ket[\vecn,- \vecn] + \frac{\sqrt{3} - e^{i \alpha}}{2\sqrt{2}} \ket[- \vecn, \vecn] \qquad (\vecn = \vecn_1, ~\vecn_2, ~ \vecn_3, ~ \vecn_4),
\end{equation}
with properly chosen directional vectors $\vecn_i$ (see \cite{tavakoliBilocalBellInequalities2021} for details), becomes the eigenbasis of $\conj (\alpha)$ exhibiting the minimum average concurrence. The eigenframe measurement in this basis is implemented in optics \cite{huangEntanglementSwappingQuantum2022}.


Now we elucidate two cases, $\alpha = \pm \pi /2$ and $\alpha = 0, \pi$, respectively presenting the maximum and the ninimum values of the concurrence \eqref{concurrence-alpha}. The concurrence takes its maximum value $4$ at $\alpha = \pm \pi /2$ where $\conj$ becomes the collective spin flip up to a constant $c$. As aforementioned, this symmetry of magnetometry has already been presented in \cite{miyazaki2022imaginarityfree}.

The concurrence takes the minimum value $2$ at $\alpha = 0, \pi$ where $\conj$ becomes
\begin{equation}
	\label{swap-flip} \swp ~ \antiu_f^{\otimes 2},
\end{equation}
up to a constant $c$. This conjugation was previously recognized as the symmetry of antiparallel spins $\ket[\vecn, - \vecn]$ \cite{miyazaki2022imaginarityfree}.
At $\alpha = 0$, the eigenframe \eqref{generalized-ejv} is the basis for the elegant joint measurement of \cite{gisinSpinFlipsQuantum1999,gisinEntanglement25Years2019}, which is known to saturate the QCRB for estimating the spin direction from antiparallel spins \cite{changetal2014}. An elegant joint measurement of this type was also used to demonstrate quantum non-locality \cite{tavakoliBilocalBellInequalities2021,pozas-kerstjensetalFullNetworkNonlocality2022} and was experimentally implemented in \cite{tangExperimentalOptimalOrienteering2020,baumerDemonstratingPowerQuantum2021,huangEntanglementSwappingQuantum2022}.

Particular GHZ-type states can be initial states satisfying the second ansatz for both the collective spin flip and the conjugation \eqref{swap-flip}. For any $K$ (and even $N$), the GHZ state-vector $\ket[GHZ_K]$ satisfies \cite{miyazaki2022imaginarityfree}
\begin{equation}
	 \antiu_f^{\otimes N} \ket[GHZ_K] = (-1)^\frac{N}{2} \ket[GHZ_K],
\end{equation}
and thus
\begin{equation}
	(\swp ~ \antiu_f^{\otimes 2})^{\otimes \frac{N}{2}} \ket[GHZ_K] = (-1)^\frac{N}{2} \ket[GHZ_K],
\end{equation}
which implies Ansatz2-S for $\conj = \antiu_f^{\otimes 2}$ and $\swp ~ \antiu_f^{\otimes 2}$, respectively. More generally, any superposed GHZ-state of the form
\begin{equation}
	\label{superposedGHZ}	\sum_{K=X,Y,Z} a_K \ket[GHZ_K], \qquad (a_X,a_Y,a_Z \in \real)
\end{equation}
also satisfies Ansatz2-S for $\conj = \antiu_f^{\otimes 2}$ and $\swp ~ \antiu_f^{\otimes 2}$. The 3D-magnetometry introduced in \cite{baumgratzdatta2016} uses a superposed GHZ-state as the initial state.

The imaginarity-free estimation of two- and three-dimensional fields, based on the superposed GHZ initial state and the eigenframe measurement of $\conj^{\otimes N/2}$, is illustrated in the circuit diagram (b) of figure \ref{fig:ifree}. A notable contrast to the scenario of one-dimensional field estimation, depicted in figure \ref{fig:ifree} (a), lies in the bilocal measurements necessary to saturate the QCRB. The underlying physics necessitating bilocal measurement is captured by Ansatz2 (G), specifically, the requirement that conjugation must anticommute with multidimensional spin operators.

In summary, we obtained the local and bilocal measurements saturating the QCRB for magnetometry by solving locality ansatzes on conjugation-symmetries. The multiqubit sensor of a one-dimensional magnetic field embraces a symmetry under local conjugation, enabling saturation of the QCRB with a local measurement. Moreover, while the multiqubit sensor of a two- and three-dimensional magnetic field cannot be made symmetric under local conjugation, it embraces a symmetry under several bilocal conjugations. The magic-basis measurement and the elegant joint measurement are particular examples of eigenframe measurements. We believe that the locality ansatzes are applicable to other sensor networks and are useful for minimizing the entanglement resources required by the measurement step.

\subsection{Example: antiparallel model}\label{subsec:antiparallel}
In \cite{miyazaki2022imaginarityfree}, one of the authors considered ``antiparallel models'' by generalizing antiparallel spins \cite{gisinSpinFlipsQuantum1999,massarCollectiveVersusLocal2000}. An antiparallel model consists of mutually conjugated state pairs and has conjugation-symmetry.

Here, let us consider antiparallel models in a QSN. Since these models have only limited applicability in practice, our interest is theoretical. As is shown below, antiparallel models rewire the entangled measurement of the network.

Let $\ket[\psi_{\bm x}]$ (${\bm x} \in X \subset \mathbb{R}^m$) be a parametrized pure quantum state on a Hilbert space $\hil$. The antiparallel model for $\ket[\psi_{\bm x}]$ consists of mutually conjugated state pairs,
\begin{equation}
 \label{antiparallel} \ket[\psi_{\bm x}] \otimes \conj \ket[\psi_{\bm x}],
\end{equation}
on $\hil^{\otimes 2}$. The antiparallel spin $\ket[\vecn, -\vecn] = \ket[\vecn] \otimes i \sigma_Y \conj \ket[\vecn]$ studied in \cite{gisinSpinFlipsQuantum1999,massarCollectiveVersusLocal2000,changetal2014} is obtained by transforming the second qubit of the antiparallel model $\ket[\vecn] \otimes \conj \ket[\vecn]$ by $i \sigma_Y$. General properties of antiparallel models are investigated in section 4.2 of \cite{miyazaki2022imaginarityfree}.
First, the antiparallel model $\{ \ketbra[\psi_{\bm x}] \otimes \conj \ketbra[\psi_{\bm x}] \conj ~|~ {\bm x} \in X \}$ is symmetric under the conjugate swap $\conj_\swp = \swp (\conj \otimes \conj)$ presented in section \ref{subsubsec:swap}. This symmetry does not depend on the original state $\ket[\psi_{\bm x}]$. Second, the quantum Fisher information of the antiparallel model $\mathcal{F}_Q^{\ket[\psi_{\bm x}] \otimes \conj \ket[\psi_{\bm x}]}$ is related to that of the original model $\mathcal{F}_Q^{\ket[\psi_{\bm x}]}$ by
\begin{equation}
	\mathcal{F}_Q^{\ket[\psi_{\bm x}] \otimes \conj \ket[\psi_{\bm x}]} ({\bm x}) = 2 \mathcal{F}_Q^{\ket[\psi_{\bm x}]} ({\bm x}), \qquad (\forall {\bm x} \in X).
\end{equation}
That is, the antiparallel model has the same quantum Fisher information as that of the original model per system size. These properties remain valid in the case of QSNs.

\begin{figure}[tb]
	\centering
	\includegraphics[width=.4\textwidth]{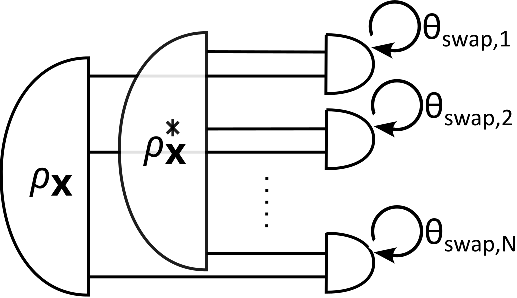}
	\caption{QCRB-saturating measurements for antiparallel models. A pair of mutually conjugated multipartite states $\ket[\psi_{\bm x}] \otimes \conj \ket[\psi_{\bm x}]$ (with a parameter ${\bm x}$) has conjugation symmetry $\swp ~ \conj \otimes \conj$. If $\conj$ is a product conjugation on the network, QCRB-saturating measurements require entanglement only between the duplicated subsystems.}
	\label{fig:antiparallel}
\end{figure}

Let us consider an antiparallel model of a multipartite state $\ket[\psi_{\bm x}]$ on
\begin{equation}
	\hil = \bigotimes_{p=1}^N \hil_p,
\end{equation}
where $\hil_p$ ($p=1,\ldots,N$) are local subsystems. As we have seen in section \ref{subsubsec:swap}, the conjugate swap is not Sep-measurable between the two subsystems, i.e., the original system $\otimes_{p=1}^N \hil_p$ and its duplicate. However, an appropriate choice of conjugation $\conj$ makes the conjugate swap Prod-measurable between doubled subsystems. A conjugation $\conj$ suitable for network sensing is the product,
\begin{equation}
	\label{localconjugate} \conj = \bigotimes_{p=1}^N \conj_p,
\end{equation}
of local conjugations $\conj_p$ on $\hil_p$ ($p=1,\ldots,N$).
\begin{theorem}
	The conjugate swap $\conj_\swp = \swp (\conj \otimes \conj)$ defined via conjugation \eqref{localconjugate} is Prod-measurable with respect to the partitioning $\hil'_1 \otimes \cdots \otimes \hil'_N$, where $\hil'_p = \hil_p \otimes \hil_p$ ($p=1,\ldots,N$).
\end{theorem}
Proof.
We observe that the conjugate swap is in a product form,
\begin{equation}
 \conj_\swp = \left( \bigotimes_{p=1}^N \swp_p \right) \times \left( \bigotimes_{p=1}^N (\conj_p \otimes \conj_p) \right) = \bigotimes_{p=1}^N \swp_p (\conj_p \otimes \conj_p),
\end{equation}
where $\swp_p:\hil_p \otimes \hil_p \rightarrow \hil_p \otimes \hil_p$ is the swap operator on $\hil_p \otimes \hil_p$. Any tensor-product of eigenframes of $\swp_p (\conj_p \otimes \conj_p)$ for $p=1,\ldots, N$ becomes an eigenframe of $\conj_\swp$.
$\qed$\\
An eigenframe measurement for an antiparallel model is presented in figure~\ref{fig:antiparallel}.
The eigenframe is entangled inside the doubled local systems $\hil_p \otimes \hil_p$ (see section \ref{subsubsec:swap} for detail), but is a product of different nodes of the network.

A notable aspect of the antiparallel model is that the eigenframe measurement remains unaffected by the original model. There might be no QCRB-saturating measurement available for the original model $\{ \ket[\psi_{\bm x}]~|~ {\bm x} \in X \}$ whatsoever. Even if it does exist, it bears no relation to the eigenframe measurement of the antiparallel model, which is entirely dictated by its symmetry.

\section{Conclusion}
\label{sec:conclusion}
We conducted an investigation into LU-equivalence classes and the LU-invariant property of measurability for multipartite conjugations. Two multipartite conjugations are considered to be LU-equivalent when it is possible to convert one into the other via a local basis transformation. A conjugation is considered to be Prod-measurable if it has product rank-1 symmetric measurements.

A conjugation is Prod-measurable if and only if it possesses a product eigenbasis. We have provided checkable necessary conditions for determining whether a conjugation is Prod-measurable. However, these conditions are generally insufficient.

For two-qubit conjugations, there exists a canonical decomposition that includes the ``magic-basis spectrum,'' which is a family of four complex unimodular numbers. The magic-basis spectrum enables us to derive the following results for two-qubit conjugations:
\begin{itemize}
\item A complete characterization of LU-equivalence classes: the space of LU-equivalence classes is homeomorphic to the configuration space of four points on a circle.
\item The LU-equivalent class of a two-qubit conjugation is invariant under subsystem permutations. No LU-invariant property can show preferences for subsystems.
\item The Prod-measurability of two-qubit conjugations can be ascertained by examining their magic-basis spectrum. The determination of the product symmetric measurement associated with a Prod-measurable conjugation can be achieved through its canonical decomposition.
\item A lower bound of entanglement resources required for implementing the conjugation-symmetric measurements can be calculated from the magic-basis spectrum.
\end{itemize}
The investigation of general multipartite conjugations is hindered by the absence of a canonical decomposition.

Conjugation symmetries have led us to QCRB-saturating measurements for several quantum sensor networks. If a pure-state model of a quantum sensor network is symmetric under a Prod-measurable conjugation, there is a product measurement that saturates the QCRB. The multiqubit networks tailored for sensing single \cite{leekokrossetastone2002,giovannettiQuantumMetrology2006,giovannettiAdvancesQuantumMetrology2011} and higher-dimensional fields \cite{baumgratzdatta2016}, along with a network rendition of the antiparallel model \cite{miyazaki2022imaginarityfree}, exhibit symmetry under product conjugations.

Our investigation of conjugations' non-locality is only preliminary, and many problems remain. Specifically, we lack a generally applicable algorithm to search for the symmetric product measurements of Prod-measurable conjugations. We believe that the most important problem is how to find conjugation symmetries for a given set of quantum states. If we had a systematic method for searching for symmetries at hand, the conjugations would lead us to optimal estimation procedures for a greater variety of quantum sensor networks.

\section*{Acknowledgements}
JM is grateful for financial support received through academist, an academic crowdfunding platform.
SA was partially supported by JST, PRESTO (Grant no. JPMJPR2111), MEXT Q-LEAP (Grant no. JPMXS0120319794), and the JST Moonshot R\&D MILLENNIA Program (Grant no. JPMJMS2061).

\bibliography{references}

\begin{thebibliography}{10}

\bibitem{stueckelbergQuantumTheoryReal1960}
Ernst C.~G. St\"{u}eckelberg.
\newblock Quantum theory in real {{Hilbert}} space.
\newblock {\em Helv. Phys. Acta}, 33(8):727--752, 1960.

\bibitem{woottersOptimalInformationTransfer2013}
William~K. Wootters.
\newblock Optimal information transfer and real-vector-space quantum theory.
\newblock In Giulio Chiribella and Robert~W. Spekkens, editors, {\em Quantum Theory: Informational Foundations and Foils}, pages 21--43. Springer Netherlands, Dordrecht, 2016.

\bibitem{casanovaQuantumSimulationMajorana2011}
J.~Casanova, C.~Sab{\'i}n, J.~Le{\'o}n, I.~L. Egusquiza, R.~Gerritsma, C.~F. Roos, J.~J. {Garc{\'i}a-Ripoll}, and E.~Solano.
\newblock Quantum {{Simulation}} of the {{Majorana Equation}} and {{Unphysical Operations}}.
\newblock {\em Phys. Rev. X}, 1(2):021018, December 2011.

\bibitem{zhangTimeReversalCharge2015}
Xiang Zhang, Yangchao Shen, Junhua Zhang, Jorge Casanova, Lucas Lamata, Enrique Solano, Man-Hong Yung, Jing-Ning Zhang, and Kihwan Kim.
\newblock Time reversal and charge conjugation in an embedding quantum simulator.
\newblock {\em Nat Commun}, 6(1):7917, November 2015.

\bibitem{dicandiaEmbeddingsimulator2013}
R.~Di~Candia, B.~Mejia, H.~Castillo, J.~S. Pedernales, J.~Casanova, and E.~Solano.
\newblock Embedding quantum simulators for quantum computation of entanglement.
\newblock {\em Phys. Rev. Lett.}, 111:240502, Dec 2013.

\bibitem{chenEfficientMeasurement2016}
Ming-Cheng Chen, Dian Wu, Zu-En Su, Xin-Dong Cai, Xi-Lin Wang, Tao Yang, Li~Li, Nai-Le Liu, Chao-Yang Lu, and Jian-Wei Pan.
\newblock Efficient measurement of multiparticle entanglement with embedding quantum simulator.
\newblock {\em Phys. Rev. Lett.}, 116:070502, Feb 2016.

\bibitem{miyazaki2022imaginarityfree}
Jisho Miyazaki and Keiji Matsumoto.
\newblock Imaginarity-free quantum multiparameter estimation.
\newblock {\em {Quantum}}, 6:665, March 2022.

\bibitem{leekokrossetastone2002}
Hwang Lee, Pieter Kok, and Jonathan~P. Dowling.
\newblock A quantum rosetta stone for interferometry.
\newblock {\em J. Mod. Opt.}, 49(14-15):2325--2338, 2002.

\bibitem{giovannettiQuantumMetrology2006}
Vittorio Giovannetti, Seth Lloyd, and Lorenzo Maccone.
\newblock Quantum metrology.
\newblock {\em Phys. Rev. Lett.}, 96(1):010401, January 2006.

\bibitem{giovannettiAdvancesQuantumMetrology2011}
Vittorio Giovannetti, Seth Lloyd, and Lorenzo Maccone.
\newblock Advances in quantum metrology.
\newblock {\em Nature Photon}, 5(4):222--229, April 2011.

\bibitem{wootters1990local}
William~K Wootters.
\newblock Local accessibility of quantum states.
\newblock {\em Complexity, entropy and the physics of information}, 8:39--46, 1990.

\bibitem{chiribellaProbabilisticTheory2010}
Giulio Chiribella, Giacomo~Mauro D'Ariano, and Paolo Perinotti.
\newblock Probabilistic theories with purification.
\newblock {\em Phys. Rev. A}, 81:062348, Jun 2010.

\bibitem{hardyLimitedHolismRealVectorSpace2012}
Lucien Hardy and William~K. Wootters.
\newblock Limited {{Holism}} and {{Real-Vector-Space Quantum Theory}}.
\newblock {\em Found Phys}, 42(3):454--473, March 2012.

\bibitem{chiribellaProcessTomography2021}
Giulio Chiribella.
\newblock Process tomography in general physical theories.
\newblock {\em Symmetry}, 13:1985, 2021.

\bibitem{wuOperationalResourceTheory2021}
Kang-Da Wu, Tulja~Varun Kondra, Swapan Rana, Carlo~Maria Scandolo, Guo-Yong Xiang, Chuan-Feng Li, Guang-Can Guo, and Alexander Streltsov.
\newblock Operational {{Resource Theory}} of {{Imaginarity}}.
\newblock {\em Phys. Rev. Lett.}, 126(9):090401, March 2021.

\bibitem{wuResourceTheoryImaginarity2021}
Kang-Da Wu, Tulja~Varun Kondra, Swapan Rana, Carlo~Maria Scandolo, Guo-Yong Xiang, Chuan-Feng Li, Guang-Can Guo, and Alexander Streltsov.
\newblock Resource theory of imaginarity: {{Quantification}} and state conversion.
\newblock {\em Phys. Rev. A}, 103(3):032401, March 2021.

\bibitem{woottersEntanglementSharingRealVectorSpace2012}
William~K. Wootters.
\newblock Entanglement {{Sharing}} in {{Real-Vector-Space Quantum Theory}}.
\newblock {\em Found Phys}, 42(1):19--28, January 2012.

\bibitem{woottersRebitThreetangleIts2014}
William~K. Wootters.
\newblock The rebit three-tangle and its relation to two-qubit entanglement.
\newblock {\em J. Phys. A: Math. Theor.}, 47(42):424037, October 2014.

\bibitem{renouQuantumTheoryBased2021a}
Marc-Olivier Renou, David Trillo, Mirjam Weilenmann, Thinh~P. Le, Armin Tavakoli, Nicolas Gisin, Antonio Ac{\'i}n, and Miguel Navascu{\'e}s.
\newblock Quantum theory based on real numbers can be experimentally falsified.
\newblock {\em Nature}, 600(7890):625--629, December 2021.

\bibitem{bednorz2022optimaldiscriminationreal}
Adam Bednorz and Josep Batle.
\newblock Optimal discrimination between real and complex quantum theories.
\newblock {\em Phys. Rev. A}, 106:042207, Oct 2022.

\bibitem{chiribellaPositiveMapsEntanglement2023}
Giulio Chiribella, Kenneth~R. Davidson, Vern~I. Paulsen, and Mizanur Rahaman.
\newblock Positive {{Maps}} and {{Entanglement}} in {{Real Hilbert Spaces}}.
\newblock {\em Ann. Henri Poincar\'e}, pages 1--30, May 2023.

\bibitem{bennettMixedStateEntanglement1996}
Charles~H. Bennett, David~P. DiVincenzo, John~A. Smolin, and William~K. Wootters.
\newblock Mixed {{State Entanglement}} and {{Quantum Error Correction}}.
\newblock {\em Phys. Rev. A}, 54(5):3824--3851, November 1996.

\bibitem{hillEntanglementPairQuantum1997a}
Scott Hill and William~K. Wootters.
\newblock Entanglement of a {{Pair}} of {{Quantum Bits}}.
\newblock {\em Phys. Rev. Lett.}, 78(26):5022--5025, June 1997.

\bibitem{woottersEntanglementFormationArbitrary1998}
William~K. Wootters.
\newblock Entanglement of {{Formation}} of an {{Arbitrary State}} of {{Two Qubits}}.
\newblock {\em Phys. Rev. Lett.}, 80(10):2245--2248, March 1998.

\bibitem{krauscirac2001}
B.~Kraus and J.~I. Cirac.
\newblock Optimal creation of entanglement using a two-qubit gate.
\newblock {\em Phys. Rev. A}, 63:062309, May 2001.

\bibitem{zhouSaturatingQuantumCramer2020}
Sisi Zhou, Chang-Ling Zou, and Liang Jiang.
\newblock Saturating the quantum {{Cram\'er}}\textendash{{Rao}} bound using {{LOCC}}.
\newblock {\em Quantum Sci. Technol.}, 5(2):025005, March 2020.

\bibitem{albarelliHolevoBound2019}
Francesco Albarelli, Jamie~F. Friel, and Animesh Datta.
\newblock Evaluating the holevo cram\'er-rao bound for multiparameter quantum metrology.
\newblock {\em Phys. Rev. Lett.}, 123:200503, Nov 2019.

\bibitem{sidhuTightBounds2021}
Jasminder~S. Sidhu, Yingkai Ouyang, Earl~T. Campbell, and Pieter Kok.
\newblock Tight bounds on the simultaneous estimation of incompatible parameters.
\newblock {\em Phys. Rev. X}, 11:011028, Feb 2021.

\bibitem{boixoQuantumlimitedMetrologyProduct2008}
Sergio Boixo, Animesh Datta, Steven~T. Flammia, Anil Shaji, Emilio Bagan, and Carlton~M. Caves.
\newblock Quantum-limited metrology with product states.
\newblock {\em Phys. Rev. A}, 77(1):012317, January 2008.

\bibitem{royExponentiallyEnhancedQuantum2008}
S.~M. Roy and Samuel~L. Braunstein.
\newblock Exponentially {{Enhanced Quantum Metrology}}.
\newblock {\em Phys. Rev. Lett.}, 100(22):220501, June 2008.

\bibitem{ramsLimitsCriticalityBasedQuantum2018}
Marek~M. Rams, Piotr Sierant, Omyoti Dutta, Pawe{\l} Horodecki, and Jakub Zakrzewski.
\newblock At the {{Limits}} of {{Criticality-Based Quantum Metrology}}: {{Apparent Super-Heisenberg Scaling Revisited}}.
\newblock {\em Phys. Rev. X}, 8(2):021022, April 2018.

\bibitem{eldredgeOptimalSecureMeasurement2018}
Zachary Eldredge, Michael {Foss-Feig}, Jonathan~A. Gross, S.~L. Rolston, and Alexey~V. Gorshkov.
\newblock Optimal and secure measurement protocols for quantum sensor networks.
\newblock {\em Phys. Rev. A}, 97(4):042337, April 2018.

\bibitem{proctorMultiparameterEstimationNetworked2018}
Timothy~J. Proctor, Paul~A. Knott, and Jacob~A. Dunningham.
\newblock Multi-parameter estimation in networked quantum sensors.
\newblock {\em Phys. Rev. Lett.}, 120(8):080501, February 2018.

\bibitem{helstromMinimumVarianceEstimates1968}
C.~Helstrom.
\newblock The minimum variance of estimates in quantum signal detection.
\newblock {\em IEEE Trans. Inf. Theory}, 14(2):234--242, Mar 1968.

\bibitem{helstrom1976}
C.~W. Helstrom.
\newblock {\em Quantum Detection and Estimation Theory}.
\newblock Academic Press, 1976.

\bibitem{buzekOptimalManipulationsQubits1999}
V.~Bu{\v z}ek, M.~Hillery, and R.~F. Werner.
\newblock Optimal manipulations with qubits: {{Universal-NOT}} gate.
\newblock {\em Phys. Rev. A}, 60(4):R2626--R2629, October 1999.

\bibitem{bunse-gerstnerSingularValueDecompositions1988}
Angelika {Bunse-Gerstner} and William~B. Gragg.
\newblock Singular value decompositions of complex symmetric matrices.
\newblock {\em J. Comput. Appl. Math.}, 21(1):41--54, January 1988.

\bibitem{chebotarevSingularValueDecomposition2014}
Alexander~M. Chebotarev and Alexander~E. Teretenkov.
\newblock Singular value decomposition for the {{Takagi}} factorization of symmetric matrices.
\newblock {\em Appl. Math. Comput.}, 234:380--384, May 2014.

\bibitem{xuDivideconquermethod2008}
Wei Xu and Sanzheng Qiao.
\newblock A divide-and-conquer method for the takagi factorization.
\newblock {\em SIAM J. Matrix Anal. Appl.}, 30(1):142--153, 2008.

\bibitem{xuTwistedfactorization2009}
Wei Xu and Sanzheng Qiao.
\newblock A twisted factorization method for symmetric svd of a complex symmetric tridiagonal matrix.
\newblock {\em Numer. Linear Algebra Appl.}, 16(10):801--815, 2009.

\bibitem{cheAdaptiveAlgorithmsComputing2018}
Maolin Che, Sanzheng Qiao, and Yimin Wei.
\newblock Adaptive algorithms for computing the principal {{Takagi}} vector of a complex symmetric matrix.
\newblock {\em Neurocomputing}, 317:79--87, November 2018.

\bibitem{zanardiEntanglingPowerQuantum2000}
Paolo Zanardi, Christof Zalka, and Lara Faoro.
\newblock Entangling power of quantum evolutions.
\newblock {\em Phys. Rev. A}, 62(3):030301, August 2000.

\bibitem{makhlin2000NonlocalPO}
Yuriy Makhlin.
\newblock Nonlocal properties of two-qubit gates and mixed states, and the optimization of quantum computations.
\newblock {\em Quant. Info. Proc.}, 1:243--252, 2000.

\bibitem{vatanOptimalQuantumCircuits2004}
Farrokh Vatan and Colin Williams.
\newblock Optimal quantum circuits for general two-qubit gates.
\newblock {\em Phys. Rev. A}, 69(3):032315, March 2004.

\bibitem{bengtssonGeometryQuantumStates2017}
Ingemar Bengtsson and Karol Zyczkowski.
\newblock {\em Geometry of {{Quantum States}}: {{An Introduction}} to {{Quantum Entanglement}}}.
\newblock {Cambridge University Press}, {Cambridge}, second edition, 2017.

\bibitem{bandyopadhyayEntanglementCostNonlocal2009}
Somshubhro Bandyopadhyay, Gilles Brassard, Shelby Kimmel, and William~K. Wootters.
\newblock Entanglement {{Cost}} of {{Nonlocal Measurements}}.
\newblock {\em Phys. Rev. A}, 80(1):012313, July 2009.

\bibitem{bandyopadhyay2010}
Somshubhro Bandyopadhyay, Ramij Rahaman, and William~K Wootters.
\newblock Entanglement cost of two-qubit orthogonal measurements.
\newblock {\em J. Phys. A: Math. Theor.}, 43(45):455303, oct 2010.

\bibitem{gisinEntanglement25Years2019}
Nicolas Gisin.
\newblock Entanglement 25 {{Years}} after {{Quantum Teleportation}}: {{Testing Joint Measurements}} in {{Quantum Networks}}.
\newblock {\em Entropy}, 21(3):325, March 2019.

\bibitem{czartowskiBipartiteQuantumMeasurements2021}
Jakub Czartowski and Karol {\.Z}yczkowski.
\newblock Bipartite quantum measurements with optimal single-sided distinguishability.
\newblock {\em Quantum}, 5:442, April 2021.

\bibitem{gisinSpinFlipsQuantum1999}
N.~Gisin and S.~Popescu.
\newblock Spin {{Flips}} and {{Quantum Information}} for {{Antiparallel Spins}}.
\newblock {\em Phys. Rev. Lett.}, 83(2):432--435, July 1999.

\bibitem{baumgratzdatta2016}
Tillmann Baumgratz and Animesh Datta.
\newblock Quantum enhanced estimation of a multidimensional field.
\newblock {\em Phys. Rev. Lett.}, 116:030801, Jan 2016.

\bibitem{holevo1982}
A.~S. Holevo.
\newblock {\em Probabilistic and Statistical Aspects of Quantum Theory}.
\newblock North Holland, 1982.

\bibitem{szczykulskaMultiparameterQuantumMetrology2016}
Magdalena Szczykulska, Tillmann Baumgratz, and Animesh Datta.
\newblock Multi-parameter quantum metrology.
\newblock {\em Adv. Phys.: X}, 1(4):621--639, Jul 2016.

\bibitem{changetal2014}
L.~Chang, N.~Li, S.~Luo, and H.~Song.
\newblock Optimal extraction of information from two spins.
\newblock {\em Phys. Rev. A}, 89:042110, Apr 2014.

\bibitem{tavakoliBilocalBellInequalities2021}
Armin Tavakoli, Nicolas Gisin, and Cyril Branciard.
\newblock Bilocal {{Bell}} inequalities violated by the quantum {{Elegant Joint Measurement}}.
\newblock {\em Phys. Rev. Lett.}, 126(22):220401, June 2021.

\bibitem{huangEntanglementSwappingQuantum2022}
Cen-Xiao Huang, Xiao-Min Hu, Yu~Guo, Chao Zhang, Bi-Heng Liu, Yun-Feng Huang, Chuan-Feng Li, Guang-Can Guo, Nicolas Gisin, Cyril Branciard, and Armin Tavakoli.
\newblock Entanglement swapping and quantum correlations via {{Elegant Joint Measurements}}.
\newblock {\em Phys. Rev. Lett.}, 129(3):030502, July 2022.

\bibitem{pozas-kerstjensetalFullNetworkNonlocality2022}
Alejandro Pozas-Kerstjens, Nicolas Gisin, and Armin Tavakoli.
\newblock Full network nonlocality.
\newblock {\em Phys. Rev. Lett.}, 128:010403, Jan 2022.

\bibitem{tangExperimentalOptimalOrienteering2020}
Jun-Feng Tang, Zhibo Hou, Jiangwei Shang, Huangjun Zhu, Guo-Yong Xiang, Chuan-Feng Li, and Guang-Can Guo.
\newblock Experimental optimal orienteering via parallel and antiparallel spins.
\newblock {\em Phys. Rev. Lett.}, 124(6):060502, February 2020.

\bibitem{baumerDemonstratingPowerQuantum2021}
Elisa B{\"a}umer, Nicolas Gisin, and Armin Tavakoli.
\newblock Demonstrating the power of quantum computers, certification of highly entangled measurements and scalable quantum nonlocality.
\newblock {\em npj Quantum Inf}, 7(1):117, July 2021.

\bibitem{massarCollectiveVersusLocal2000}
S.~Massar.
\newblock Collective versus local measurements on two parallel or antiparallel spins.
\newblock {\em Phys. Rev. A}, 62:040101, Sep 2000.

\bibitem{hong1986213}
Yoo~Pyo Hong, Roger~A. Horn, and Charles~R. Johnson.
\newblock On the reduction of pairs of hermitian or symmetric matrices to diagonal form by congruence.
\newblock {\em Linear Algebra Appl.}, 73:213--226, 1986.

\bibitem{wignerNormalFormAntiunitary1960}
Eugene~P. Wigner.
\newblock Normal {{Form}} of {{Antiunitary Operators}}.
\newblock {\em J. Math. Phys.}, 1(5):409--413, September 1960.

\end{thebibliography}

\appendix

\section{Local measurability}
\label{appx:Prod-measurability}
Here, we provide proofs of the theorems and corollaries presented in section~\ref{sec:Prod-measurable}.

\subsection{Product eigenbases of Prod-measurable conjugations (theorem \ref{thm:prod})}
\label{appx:proof_prod}
\begin{definition}[connected vectors]
	A pair of vectors $\ket[\psi_\mathrm{ini}]$ and $\ket[\psi_\mathrm{fin}]$ in a set of vectors $\Psi$ is said to be \emph{connected} if there is a sequence of elements $\ket[\psi_j] \in \psi$ ($j=1,\ldots,n$) such that $\ket[\psi_1] = \ket[\psi_\mathrm{ini}]$, $\ket[\psi_n] = \ket[\psi_\mathrm{fin}]$ and
	\begin{equation}
 		\braket{\psi_j}{\psi_{j+1}} \neq 0 ~~(j=1,\ldots,n-1).
	\end{equation}
	A subset of pairwise connected elements in a vector set is called a \emph{connected component}.
\end{definition}
From here, we consider an $N$-partite system $\otimes_{p=1}^N \hil_p$ and a conjugation $\conj$ thereon.
\begin{lemma}\label{lem:connected}
Let $\Psi_p \subset \hil_p$ $(p=1,\ldots,N)$ be a set of vectors such that there exist phases $\phi_{\psi^1 \ldots \psi^N}$ $(\ket[\psi^p] \in \Psi_p,~p=1,\ldots,N)$ satisfying

 \begin{equation}
 \label{eigenphase} \conj \ket[\psi^1,\ldots,\psi^N] = e^{i \phi_{\psi^1 \ldots \psi^N}} \ket[\psi^1,\ldots,\psi^N].
 \end{equation}
If all the vectors in $\Psi_p$ are connected, there are a maximal real subspace $\R_p$ of $\mathrm{span}_\mathbb{C} \langle \Psi_p \rangle$, and $\psi^p$-independent phases $\phi_{\psi^1 \ldots \psi^{p-1} \psi^{p+1} \ldots \psi^N}$ such that 
 \begin{equation}
 \label{replace} \conj \ket[\psi^1, \ldots, \psi^{p-1}, \eta, \psi^{p+1}, \ldots, \psi^N] = e^{i \phi_{\psi^1 \ldots \psi^{p-1} \psi^{p+1} \ldots \psi^N}} \ket[\psi^1, \ldots, \psi^{p-1}, \eta, \psi^{p+1}, \ldots, \psi^N].
 \end{equation}
holds for all combinations of vectors $\ket[\eta] \in \R_p$ and $\ket[\psi^p] \in \Psi_p$ ($p=1,\ldots,p-1,p+1,\ldots,N$).
\end{lemma}
Proof.
We assume $p=1$ for simplicity of notation. For any pair of vectors $\ket[\psi^1],~ \ket[\psi'^1] \in \Psi_1$ and $\ket[\psi^2,\ldots,\psi^N] \in \otimes_{p=2}^N \Psi_p$, we have
\begin{align}
 e^{i \phi_{\psi'^1 \psi^2 \ldots \psi^N}} \braket{\psi^1}{\psi'^1} &= \bra[\psi^1,\psi^2,\ldots,\psi^N] \conj \ket[\psi'^1,\psi^2,\ldots,\psi^N] \\
 &= \bra[\psi'^1,\psi^2,\ldots,\psi^N] \conj \ket[\psi^1,\psi^2,\ldots,\psi^N] \\
 &= e^{i \phi_{\psi^1 \psi^2 \ldots \psi^N}} \braket{\psi'^1}{\psi^1},
\end{align}
where the second line follows from the definition of the Hermitian adjoint for antilinear operators. If $\ket[\psi]$ and $\ket[\psi']$ are not orthogonal, this results in 
\begin{equation}
\label{transition} e^{i \phi_{\psi'^1 \psi^2 \ldots \psi^N}} = e^{i \phi_{\psi^1 \psi^2 \ldots \psi^N}} c_{\psi^1 \rightarrow \psi'^1} \qquad \left( c_{\psi^1 \rightarrow \psi'^1} := \frac{\braket{\psi'^1}{\psi^1}}{\braket{\psi^1}{\psi'^1}} \right).
\end{equation}

Let us choose a vector $\ket[\psi^1_\mathrm{ini}]$ from the connected component $\Psi_1$. Since $\ket[\psi^1_\mathrm{ini}]$ and any vector $\ket[\psi^1] \in \Psi_1$ are connected through a sequence $\ket[\psi^1_j]$ ($j=1,\ldots,n$), \eqref{transition} implies
\begin{equation}
\label{factorizing_phase} e^{i \phi_{\psi^1 \psi^2 \ldots \psi^N}} = e^{i \phi_{\psi^1_\mathrm{ini} \psi^2 \ldots \psi^N}} \times e^{i \phi_{\psi^1}},
\end{equation}
for any $\ket[\psi^p] \in \Psi_p, ~p=1,\ldots,N$, where
\begin{equation}
 \label{pathind_phase} \phi_{\psi^1} = \arg \left[ c_{\psi^1_\mathrm{ini} \rightarrow \psi^1_2} \times \ldots \times c_{\psi^1_{n-1} \rightarrow \psi^1} \right].
\end{equation}
Relation \eqref{factorizing_phase} reveals that $\phi_{\psi^1}$ does not depend on the connected sequence from $\ket[\psi^1_\mathrm{ini}]$ to $\ket[\psi^1]$. Therefore, the phases \eqref{pathind_phase} satisfy
\begin{align}
 \psi_{\phi^1} - \psi_{\phi'1} &= \arg \left[ c_{\psi^1_\mathrm{ini} \rightarrow \psi^1_2} \times \ldots \times c_{\psi^1_{n-1} \rightarrow \psi^1} \right] - \arg \left[ c_{\psi^1_\mathrm{ini} \rightarrow \psi^1_2} \times \ldots \times c_{\psi^1_{n-1} \rightarrow \psi^1} \times c_{\psi^1 \rightarrow \psi'^1} \right] \\
 \label{phase_difference} &= \arg \left[ c_{\psi^1 \rightarrow \psi'^1} \right] = \arg \left[ \frac{\braket{\psi'^1}{\psi^1}}{\braket{\psi^1}{\psi'^1}} \right]
\end{align}
if $\ket[\psi^1]$ and $\ket[\psi'^1]$ are non-orthogonal elements in $\Psi_1$.

Now, let us construct the desired maximal real Hilbert subspace and the $\psi^1$-independent phases. Adjust the phases of each $\ket[\psi^1] \in \Psi_1$ to define vectors,
\begin{equation}
 \ket[r(\psi^1)] := e^{i \frac{\phi_{\psi^1}}{2}} \ket[\psi^1].
\end{equation}
These newly defined vectors live in a real Hilbert space since their inner products are all real numbers. In fact, \eqref{phase_difference} implies
\begin{equation}
 \braket{r(\psi^1)}{r(\psi'^1)} = e^{i (\phi_{\psi^1}- \phi_{\psi'^1})/2} \braket{\psi^1}{\psi'^1} 
 = \sqrt{\frac{\braket{\psi'^1}{\psi^1}}{\braket{\psi^1}{\psi'^1}}} \braket{\psi^1}{\psi'^1} = |\braket{\psi'^1}{\psi^1}|,
\end{equation}
if $\ket[\psi^1]$ and $\ket[\psi'^1]$ are non-orthogonal elements in $\Psi_1$. The real subspace
\begin{equation}
 \R = \mathrm{span}_\real \langle r(\psi^1) ~|~ \ket[\psi^1] \in \Psi_1 \rangle
\end{equation}
of $\hil_1$ is maximal since $\ket[r(\psi^1)]$ and $\ket[\psi^1]$ differs only in their phases, and since $\Psi_1$ spans $\hil_1$ through its complex coefficients.

Define a $\psi^1$-independent phase $\phi_{\psi^2 \ldots \psi^N}$ by
\begin{equation}
 \phi_{\psi^2 \ldots \psi^N} := \phi_{\psi^1_\mathrm{ini} \psi^2 \ldots \psi^N},
\end{equation}
for any combination of $\ket[\psi^p] \in \Psi_p$ $(p \neq 1)$. We have
\begin{equation}
 \conj \ket[r(\psi^1),\psi^2,\ldots,\psi^N] = e^{i \phi_{\psi^2 \ldots \psi^N}} \ket[r(\psi^1),\psi^2,\ldots,\psi^N],
\end{equation}
for any $\ket[\psi^1] \in \Psi_1$, and thus,
\begin{equation}
 \conj \ket[\eta,\psi^2,\ldots,\psi^N] = e^{i \phi_{\psi^2 \ldots \psi^N}} \ket[\eta,\psi^2,\ldots,\psi^N],
\end{equation}
for any vector $\ket[\eta] \in \R$.
$\qed$

Proof of theorem \ref{thm:prod}.
Let $\conj$ be a Prod-measurable conjugation and $\{ \ket[\psi^1_{j_1},\ldots,\psi^N_{j_N}] \}_{j_p \in J_p, p=1,\ldots,N}$ be its product eigenframe on $\otimes_{p=1}^N \hil_p$. The product eigenframe must satisfy
\begin{equation}
 \conj \ket[\psi^1_{j_1},\ldots,\psi^N_{j_N}] = e^{i \phi_{j_1 , \ldots ,j_N}} \ket[\psi^1_{j_1},\ldots,\psi^N_{j_N}],
\end{equation}
with some phase $\phi_{j_1 , \ldots ,j_N}$.

Separate $\{ \ket[\psi^1_{j_1}] \}_{j_1 \in J_1}$ into connected components, say $\Psi^1_k$ ($k=1,\dots,K$). By applying lemma~\ref{lem:connected} to all the components, we find the maximal real subspaces $\R^1_k$ of $\mathrm{span}_{\mathbb{C}} \Psi^1_k$, each satisfying \eqref{replace}. These real subspaces are orthogonal to each other since the connected components $\Psi^1_k$ are mutually disconnected. Take a basis $\{ \ket[\eta_j] \}_{j=1,\ldots,\dim \hil_1}$ of $\hil_1$ so that each element $\ket[\eta_j]$ is in either of the real subspace $\R^1_k$. From \eqref{replace}, for any $j=1,\ldots,\dim \hil_1$, we have
\begin{equation}
 \conj \ket[\eta_j, \psi^2_{j_2},\ldots,\psi^N_{j_N}] = e^{i \phi_{j, \psi^2_{j_2},\ldots,\psi^N_{j_N}}} \ket[\eta_j, \psi^2_{j_2},\ldots,\psi^N_{j_N}],
\end{equation}
with some phase $\phi_{j, \psi^2_{j_2},\ldots,\psi^N_{j_N}}$. (The phase $\phi_{j, \psi^2_{j_2},\ldots,\psi^N_{j_N}}$ depends on $j$ only through $k$, the index of the real subspace $\R^1_k$ containing $\ket[\eta_j]$.)

By repeating the replacement of $\{ \ket[\psi^p_{j_p}] \}_{j_p \in J_p}$ with $\{ \ket[\eta^p_j] \}_{j=1,\ldots,\dim \hil_p}$ for all subsystems, we arrive at the product basis $\{ \ket[\eta^1_{j_1}, \ldots, \eta^N_{j_N} ] \}_{j_p=1,\ldots,\dim \hil_p, ~ p=1,\ldots,N}$ satisfying
\begin{equation}
 \conj \ket[\eta^1_j, \ldots,\eta^N_{j_N}] = e^{i \phi_{j_1, \ldots, j_N}} \ket[\eta^1_j, \ldots,\eta^N_{j_N}],
\end{equation}
with some phase $\phi_{j_1, \ldots, j_N}$.
$\qed$

\subsection{Conditions for Prod-measurability}
\label{appx:proof_search}
\subsubsection*{Proof of corollary \ref{cor:search}}
Let $\{ \ket[\psi^p_j] \}_{j=1,\ldots,\dim \hil_p}$ be the $p$-th part of the basis that diagonalizes $[\conj]$ by \eqref{product_diagonal}, for $p=1,\ldots ,N$. We construct the Takagi matrices $V^p$ of $\trace_{\overline{p}} \left[ [\conj]^\compt \right]$ from the unitary matrices $U^p$ of the basis transformations,
\begin{equation}
 \label{local_basis_transformation} U^p_{j_p k_p} := \braket{\psi^p_{k_p}}{j_p}.
\end{equation}
We have
\begin{equation}
 U [\conj]^\compt U^\top = [\conj]^\psi, \qquad \left( U := \bigotimes_{p=1}^N U^p \right).
\end{equation}
A direct calculation yields
\begin{align}
 \label{diagonal_traces} \trace_{\overline{p}} \left[ [\conj]^\compt \right] &= \trace_{\overline{p}} \left[ U^\dagger [\conj]^\psi U^\ast \right] = U^{p \dagger} \mathrm{diag} (c_1,\ldots,c_{\dim \hil_p}) U^{p \ast} \\
 &= V^{p \dagger} \mathrm{diag} (|c_1|,\ldots,|c_{\dim \hil_p}|) V^{p \ast}, \\
 \label{c} & \left( c_{j_p} := \sum_{j_1,\ldots,j_{p-1},j_{p+1},\ldots,j_N} e^{i \phi_{j_1 , \ldots ,j_N}} \prod_{q \neq p} [U^{q \dagger} U^{q \ast}]_{j_q j_q} \right)
\end{align}
where $V^p$ is defined by
\begin{equation}
 V^p = D_p U^p, \qquad \left( \mathrm{diag} \left( e^{- i \arg (c_1) /2}, \ldots, e^{- i \arg (c_{\dim \hil_p}) /2} \right) \right)
\end{equation}
for $p=1,\ldots,N$.

$V^p$ are Takagi matrices of $\trace_{\overline{p}} \left[ [\conj]^\compt \right]$ with the desired property. The product $V := \otimes_p V^p$ diagonalizes $[\conj]^\compt$ into
\begin{equation}
 V [\conj]^\compt V^\top = \otimes_{p=1}^N D_p [\conj]^\psi \otimes_{p=1}^N D_p,
\end{equation}
where both $\otimes_{p=1}^N D_p$ and $[\conj]^\psi$ are diagonal.

\subsubsection{Proof of corollary \ref{cor:necessary} (total-normality criterion)}
From corollary~\ref{cor:search}, the partial traces $\trace_{\overline{p}} \left[ [\conj]^\compt \right]$ of a Prod-measurable conjugation $\conj$ can be diagonalized by performing congruence transformations for all $p$. Therefore, $\trace_{\overline{p}} \left[ [\conj]^\compt \right]$ must be symmetric for all $p$.

From corollary~\ref{cor:search}, if $\conj$ is Prod-measurable, $[\conj]^\compt$ and $\otimes_{p=1}^N \trace_{\overline{p}} \left[ [\conj]^\compt \right]$ are diagonalized by the same unitary congruence transformation. A non-singular symmetric matrix $A$ and a general symmetric matrix $B$ are diagonalized by the same unitary congruence transformation if and only if $A^{-1}B$ is normal \cite{hong1986213}. The corollary follows since $[\conj]^\compt$ is non-singular and the inverse matrix is $\left( [\conj]^\compt \right)^\dagger$.

\subsubsection{Proof of corollary \ref{cor:sufficient}}
Let $V^p$ be a Takagi matrix of $\trace_{\overline{p}} \left[ [\conj]^\compt \right]$ for $p=1,\ldots,N$. Assuming that the Takagi values are non-degenerate, $\otimes_p V^p$ is the unique Takagi matrix of $\otimes_{p=1}^N \trace_{\overline{p}} \left[ [\conj]^\compt \right]$ up to permutation of the column vectors.

Let $U$ be the unitary matrix that simultaneously diagonalizes $[\conj]^\compt$ and $\otimes_{p=1}^N \trace_{\overline{p}} \left[ [\conj]^\compt \right]$ by the congruence transformation. If $U$ diagonalizes $\otimes_{p=1}^N \trace_{\overline{p}} \left[ [\conj]^\compt \right]$ to
\begin{equation}
 U \otimes_{p=1}^N \trace_{\overline{p}} \left[ [\conj]^\compt \right] U^\top = \mathrm{diag} (r_1 e^{i \phi_1}, \ldots, r_{\dim \hil} e^{i \phi_{\dim \hil}}), \qquad \left( r_j \geq 0 \right),
\end{equation}
then the $\dim \hil \times \dim \hil$ matrix $V$ defined by
\begin{equation}
 V = \mathrm{diag} (e^{-i \phi_1/2}, \ldots, e^{i \phi_{\dim \hil}/2}) U,
\end{equation}
is a Takagi matrix of $\otimes_{p=1}^N \trace_{\overline{p}} \left[ [\conj]^\compt \right]$. Since $U$ diagonalizes $[\conj]^\compt$, so does $V$. 

The Takagi matrices $\otimes_p V^p$ and $V$ of $\otimes_{p=1}^N \trace_{\overline{p}} \left[ [\conj]^\compt \right]$ are mutually related by a permutation of column vectors, since the Takagi values are non-degenerate. This implies that $\otimes_p V^p$ diagonalizes $[\conj]^\compt$ as well as $V$. Conjugation $\conj$ is Prod-measurable since $[\conj]^\compt$ is diagonalized by a product unitary matrix.

\subsection{Proof of theorem \ref{thm:prod}}
\label{appx:poof_prod_meas_qubit}
Let $\conj$ be a Prod-measurable two-qubit conjugation. By theorem \ref{thm:prod}, $\conj$ has a product eigenframe $\psi = \{ \ket[\psi^A_{j_A}, \psi^B_{j_B}] \}_{j_A,j_B = 0,1}$ satisfying $\conj \ket[\psi^A_{j_A}, \psi^B_{j_B}] = e^{i \phi_{j_A,j_B}} \ket[\psi^A_{j_A}, \psi^B_{j_B}]$. The magic-basis spectrum of this conjugation is given by $\{ e^{i (\phi_{0,0} + \phi_{1,1})}, - e^{i (\phi_{0,0} + \phi_{1,1})}, e^{i (\phi_{0,1} + \phi_{1,0})}, -e^{i (\phi_{0,1} + \phi_{1,0})} \}$. While the phases $\phi_{j_A,j_B}$ can take any values, the magic-basis spectrum of Prod-measurable two-qubit conjugation is equivalent to $\{ 1,-1,z,-z \}$ with some unimodular $z$.

Since Prod-measurability is an LU-invariant property, if a magic-basis spectrum of a conjugation is equivalent to $\{ 1,-1,z,-z \}$ with some unimodular $z$, then the conjugation is Prod-measurable.

All product conjugations are LU-equivalent to each other. The complex conjugation in the two-qubit computational basis is a product conjugation and has the magic-basis spectrum $\{ 1,-1,1,-1 \}$. Therefore, the magic-basis spectra of product conjugations are equivalent to $\{ 1,-1,1,-1 \}$.

\section{Conjugations made of antiunitaries}
\label{sec:conjugationfromantiunitary}
The collective spin flip $\antiu_f^{\otimes 2}$ is a conjugation, while its components $\antiu_f$ are not. We show a general form of antiunitary pairs that compose into a conjugation.

For this purpose, Wigner's canonical form of antiunitaries is useful. Matrix representations of antiunitary operators are unitary matrices. A unitary matrix is diagonalized by a unitary congruence transformation if and only if it is symmetric, that is, if it represents a conjugation. In general, it is possible to turn any unitary matrix into a Wigner canonical form \cite{wignerNormalFormAntiunitary1960}
\begin{equation}
 \left(
 \begin{array}{cccc}
 1_{d-(m/2)} &&& \\
 & A_{\omega_1} && \\
 && \ddots & \\
 &&& A_{\omega_m}
 \end{array}\right), \qquad
 A_\omega = \left(
 \begin{array}{cc}
 0 & 1 \\
 \omega & 0
 \end{array}\right),
\end{equation}
through a suitable unitary congruence transformation. Here, $1_{d-(m/2)}$ is a $d-(m/2)$-dimensional identity matrix, and $\omega \neq 1$ is some unimodular number.

\begin{theorem}\label{thm:onlyflip}
Let $\antiu_1$ and $\antiu_2$ be antiunitaries that are not conjugations. Then $\antiu_1 \otimes \antiu_2$ is a conjugation if and only if $\antiu_1$ and $\antiu_2$ are unitarily equivalent to direct sums of spin flips.
\end{theorem}
Proof.
Let
\begin{align}
 U_j = \mathbb{I}_{\dim \hil_j - (m_j/2)} \bigoplus_{k=1}^{m_j} A_{\omega^j_k}, \qquad ( \omega^j_k \neq 1 ),
\end{align}
be the canonical form of $[\antiu_j]$ for $j=1,2$. The tensor product $\antiu_1 \otimes \antiu_2$ is a conjugation if and only if
\begin{align}
 U_1 \otimes U_2 &= A \oplus B \oplus C \\
 A &= \mathbb{I}_{\dim \hil_1 - (m_1/2)} \otimes \left( \oplus_{k=1}^{m_2} A_{\omega^2_k} \right) \\
 B &= \left( \oplus_{k=1}^{m_1} A_{\omega^1_k} \right) \otimes \mathbb{I}_{\dim \hil_2 - (m_2/2)} \\
 C &= \left( \oplus_{k=1}^{m_1} A_{\omega^1_k} \right) \otimes \left( \oplus_{k=1}^{m_2} A_{\omega^2_k} \right),
\end{align}
is symmetric in the sense that $U_1^\top \otimes U_2^\top = A^\top \oplus B^\top \oplus C^\top = A \oplus B \oplus C$. $A^\top = A$ and $B^\top = B$ imply $\dim \hil_j - (m_j/2)= 0$ for $j=1,~2$. $C^\top =C$ implies
\begin{align}
 \left( \begin{array}{cc} 0 & 1 \\ \omega^1_k & 0 \end{array} \right)^\top \otimes \left( \begin{array}{cc} 0 & 1 \\ \omega^2_{k'} & 0 \end{array} \right)^\top = \left( \begin{array}{cc} 0 & 1 \\ \omega^1_k & 0 \end{array} \right) \otimes \left( \begin{array}{cc} 0 & 1 \\ \omega^2_{k'} & 0 \end{array} \right) \\ \Leftrightarrow \omega^1_k = \omega^2_{k'} ,~\omega^1_k \omega^2_{k'} =1 \Leftrightarrow \omega^1_k = \omega^2_{k'} = -1,
\end{align}
for any $k= 1, \ldots ,m_1$ and $k'= 1, \ldots,m_2$. Therefore, $U_j$ are direct sums of $A_{-1}$ blocks, implying that $\antiu_j$ are direct sums of spin flips for $j=1,2$.
$\qed$\\

\begin{corollary}\label{cor:onlytwo}
Let $\antiu_p$ be antiunitary operators on $\hil_p$ for $p=1,\ldots,N$. If (1) $\otimes_{p=1}^N \antiu_p$ is a conjugation, and if (2) $\otimes_{p \in X} \antiu_p$ is not a conjugation for any proper subset $X \subset \{ 1,\ldots, N \}$, then $N=2$.
\end{corollary}
Proof.
Let $\antiu_p$ ($p=1,\ldots,N$) be antiunitaries satisfying the assumptions in corollary~\ref{cor:onlytwo}. For any $p$, $\antiu_p$ is not a conjugation from the first assumption. Since $\antiu_p \otimes \bigotimes_{p' \neq p} \antiu_{p'}$ is a conjugation, $\antiu_p$ is unitarily equivalent to a direct sum of spin flips for any $p$. For any pair of subsystems $p$ and $p'$, $\antiu_p \otimes \antiu_{p'}$ is a conjugation. We have $N=2$ from the second assumption.
$\qed$\\

\section{Entanglement in eigenframes of two-qubit conjugations}
\label{appx:entanglement_two-qubit}
Recall the properties of concurrences \cite{hillEntanglementPairQuantum1997a,woottersEntanglementFormationArbitrary1998}. Let $\ket[\psi]$ be a unit vector on a two-qubit system. The concurrence $C(\psi)$ of a pure state $\ket[\psi]$ is
\begin{equation}
 C(\psi) := \frac{| \langle \psi | \widetilde{\psi} \rangle |}{\langle \psi | \psi \rangle}, \qquad \left( \ket[ \widetilde{\psi} ] := \sigma_Y \otimes \sigma_Y \ket[\psi^\ast] \right),
\end{equation}
and is related to the entanglement entropy $E(\psi)$ by
\begin{equation}
 E(\psi) = h \left( \frac{1+\sqrt{1-C(\psi)^2}}{2} \right), \qquad \left(h(p)= - p \log p - (1-p) \log (1-p) \right).
\end{equation}
Let $\mathcal{V} = \{ \ket[v_j] \}_{j=1,\ldots,n}$ be a frame. We define the maximum and average concurrences and the maximum and average entanglements of $\mathcal{V}$ by
\begin{align}
		C_{max} (\mathcal{V}) &:= \max_j C(v_j), \hspace{1cm} C_{ave} (\mathcal{V}) := \sum_{j=1}^n \langle v_j | v_j \rangle C(v_j)\\
		E_{max} (\mathcal{V}) &:= \max_j E(v_j), \hspace{1cm} E_{ave} (\mathcal{V}) := \sum_{j=1}^n \langle v_j | v_j \rangle E(v_j).
\end{align}
Note that $\langle v_j | v_j \rangle$ is not a normalized probability distribution. For $2 \times 2 = 4$ dimensional systems, $\sum_j \langle v_j | v_j \rangle = \dim \hil = 4$.

All eigenframes of a qubit-qubit conjugations are generated from a single eigenbasis by orthogonal transformations. Fix an eigenbasis of the qubit-qubit conjugation $\conj$ with real eigenvalues (that is, $+1$ or $-1$), and denote it by $\{ \ket[x_k] \}_{k=0,1,2,3}$. Any eigenvector of $\conj$ with a real eigenvalue is represented by a real sum $\sum_{k=1}^4 q_k \ket[x_k]$. A set of $n$ vectors $\{ \ket[y_j] := \sum_{k=1}^4 R_{jk} \ket[x_k] \}_{j=0,\ldots,n-1}$ satisfies $\sum_{j=1}^n \ketbra[y_j] = \id_4$ if and only if the real $n \times 4$ matrix $R$ satisfies $[R^\top R]_{kl} = \sum_{j=1}^n R_{jk} R_{jl} = \delta_{kl}$. In summary, for any eigenframe $\mathcal{Y}=\{ \ket[y_j] \}_{j=0,\ldots,n-1}$, there is an $n \times 4$ real matrix $R$ such that
\begin{align}
	\label{orthogonal}	R^\top R &= \id_4 \\
	\label{eigenframe_y}	\ket[y_j] &= \sum_{k=1}^4 R_{jk} \ket[x_k], \qquad j=0,\ldots,n-1,
\end{align}
and conversely, any real $n \times 4$ matrix $R$ satisfying \eqref{orthogonal} generates an eigenframe by \eqref{eigenframe_y}. The average fidelity reads
\begin{align}
	C_{ave} (\mathcal{Y}) &= \sum_{j=1}^n \langle y_j | y_j \rangle C(y_j) \\
	C(y_j) &= \frac{| \langle y_j | \widetilde{y_j} \rangle |}{\langle y_j | y_j \rangle} = \frac{\left| \sum_{k,l=0}^3 R_{jk} R_{jl} \langle x_k | \widetilde{x_l} \rangle \right|}{ \sum_{k=1}^4 R_{jk} R_{jk} } = \frac{\left| [RWR^\top]_{jj} \right|}{[RR^\top]_{jj}},
\end{align}
where
\begin{equation}
	W_{kl} := \langle x_k | \widetilde{x_l} \rangle .
\end{equation}

\subsection{Proof of theorem \ref{thm:concurrence}}
Let $\{ e^{i \phi_j} \}_{j=1,\ldots,4 }$ and $\magic = \{ \ket[\magic_j] \}_{j=1,\ldots,4 }$ be as defined in the Theorem. Namely, they are the magic-basis spectrum and corresponding canonical magic basis of the two-qubit conjugation $\conj$, such that the eigenequations $\conj \ket[\magic_j] = e^{i \phi_j} \ket[\magic_j]$ hold for $p=1,\ldots,4$.

The vectors
\begin{equation}
 \ket[x_j] := e^{i \phi_j /2} \ket[\mu_j], \qquad j=1,\ldots,4
\end{equation}
form a reference basis of $\conj$. The matrix $W = [ \braket{x_k}{\tilde{x_l}} ]$ for this reference basis can be diagonalized to $W= - \mathrm{diag}(e^{-i \phi_1}, e^{-i \phi_2}, e^{-i \phi_3}, e^{-i \phi_4})$. The average concurrence of the eigenframe $\mathcal{Y}=\{ \ket[y_j] \}_{j=1,\ldots,n}$, defined by \eqref{eigenframe_y}, reads
\begin{equation}
 \label{lowerbound_average} C_{ave}(\mathcal{Y}) = \sum_{j=1}^n \left| \sum_{k=1}^4 e^{- i \phi_k} R_{jk}^2 \right| \geq \left| \sum_{j=1}^n \sum_{k=1}^4 e^{- i \phi_k} R_{jk}^2 \right| = \left| \sum_{k=1}^4 e^{- i \phi_k} \right|,
\end{equation}
where the last equality follows from condition \eqref{orthogonal} on the orthogonal matrix.

Let us choose a real $4 \times 4$ Hadamard matrix, divided by $2$, as $R$. The resulting reference basis $\mathcal{V} = \{ v_j \}_{j=1,\ldots,4}$, explicitly defined by \eqref{Hadamard}, has the average concurrence of
\begin{equation}
 C_{ave}(\mathcal{V}) = \sum_{j=1}^n \left| \sum_{k=1}^4 e^{- i \phi_k} \frac{H_{jk}^2}{4} \right| = \frac{1}{4} \sum_{j=1}^n \left| \sum_{k=1}^4 e^{- i \phi_k} \right| = \left| \sum_{k=1}^4 e^{- i \phi_k} \right|.
\end{equation}
Since this equals the lower bound \eqref{lowerbound_average}, the lower bound is saturated by the eigenframe $\mathcal{V}$.

\subsection{Proof of theorem \ref{thm:no_sep}}
\label{appx:proof_product_qubit}
A two-qubit conjugation is Sep-measurable if and only if its minimum average concurrence \eqref{minimum_concurrence} is zero. Here, we demonstrate that the occurrence of zero average concurrence is exclusive to cases where the magic-basis spectrum is equivalent to $\{ 1,-1, e^{i \phi}, -e^{i \phi} \}$ with an arbitrary phase $\phi$.

According to theorem \ref{thm:concurrence}, the minimum average concurrence \eqref{minimum_concurrence} becomes zero precisely when 
\begin{equation}
 \label{sum-1} \sum_{j=1}^4 e^{i \phi_j} = 0.
\end{equation}
To visualize this condition, envision a trajectory on a complex plane with four directed edges, denoted as ${\bm r}_1,\ldots,~ {\bm r}_4$, originating from the origin. The vector components of ${\bm r}_j$ are assigned as $(\cos \phi_j, ~\sin \phi_j )$ for $j=1,\ldots,4$ to ensure that the trajectory terminates at the point $\sum_{j=1}^4 e^{i \phi_j}$. Condition \eqref{sum-1} indicates that the trajectory must end at the origin. As each edge has unit length, the four edges must consist of two pairs of mutually antiparallel edges. Consequently, the magic-basis spectrum must be equivalent to $\{ e^{i \phi'},~-e^{i \phi'},~e^{i \phi''},~e^{-i \phi''} \}$ with arbitrary phases $\phi'$ and $\phi''$, and thus to $\{ 1,-1,~e^{i (\phi''-\phi')},~e^{-i (\phi''-\phi')} \}$.

In conclusion, the minimum average concurrence \eqref{minimum_concurrence} of a two-qubit conjugation reaches zero if and only if its magic-basis spectrum is equivalent to $\{ 1,-1,~e^{i \phi},~e^{-i \phi} \}$ with an arbitrary phase $\phi$. As per theorem \ref{thm:product_qubit}, such conjugations are Prod-measurable.

\subsection{Eigenframes of conjugate swap with minimum entanglement}
\label{appx:ejm-minimum}
The elegant joint measurement \eqref{ejv_2} exhibits the minimum average concurrence among all the eigenframes of conjugate swap. Here, we prove a stronger statement: it is the unique measurement that minimizes the maximum concurrence $C_{max}$.

\begin{theorem}\label{thm:ejm}
Among the eigenframes of two-qubit $\conj_\swp$, the minimum of $C_{max}$ is attained only by an eigenframe $\mathcal{V}$ with the following property: there are real numbers $r_j$ and directional vectors $\vecn_j$ ($j=1,\ldots,m$) such that
	\begin{align}
		\label{EJV}	& \ketbra[v_j] = r_j^2 \ketbra[\Psi_{\vecn_j}] \qquad j=1,\ldots,m,\\
		\label{zerosum}	& \sum_{j=1}^m r_j^2 = 4, \qquad \sum_{j=1}^m r_j^2 \vecn_j = \mathbf{0}, \qquad \sum_{j=1}^m r_j^2 n^\eta_j n^\mu_j = \frac{4}{3} \delta_{\eta \mu}, \qquad \eta,~\mu=x,~y,~z,
	\end{align}
where $\ket[\Psi_\vecn]$ is the vector defined by \eqref{ejv_2}. The eigenframe $\mathcal{V}$ also attains the minimum of $C_{ave}$ for all eigenframes of $\conj_\swp$. In particular, when $m=4$, $\mathcal{V}$ represents an elegant joint measurement with directional vectors pointing at the vertices of a tetrahedron.
\end{theorem}
Proof.
By definition, we have
\begin{equation}
 4 C_{max}(\mathcal{W}) \geq C_{ave} (\mathcal{W}),
\end{equation}
for a general frame $\mathcal{W} = \{ \ket[w_j] \}_{j=1,\ldots,m}$, with equality holding if and only if the frame vectors possess identical concurrences $C(w_j) = C_{max}(\mathcal{W}) = C_{ave}(\mathcal{W})/4$ ($j=1,\ldots,m$). We will establish that the eigenframe $\mathcal{V}$ represents the sole solution that minimizes $C_{ave}$ with the frame vectors with identical concurrences. Hence, it is the unique eigenframe that minimizes $C_{max}$.

All eigenframes of $\conj_\swp$ are generated from a single eigenbasis by orthogonal transformations, as in \eqref{eigenframe_y}. We choose a specific reference basis,
\begin{equation}
	\label{reference}	\ket[x_0] := \ket[\Psi_+], \qquad \ket[x_1] := \ket[\Psi_-], \qquad \ket[x_2] := \ket[\Phi_+], \qquad \ket[x_3] := i \ket[\Phi_-],
\end{equation}
of $\conj_\swp$. The matrix $W$ reduces to the diagonal form $\mathrm{diag}(-1,1,1,1)$.

The concurrence for each eigenvector $\ket[y_j]$ reduces to
\begin{equation}
	C(y_j) = \frac{| - R_{j0}^2 + R_{j1}^2 + R_{j2}^2 + R_{j3}^2 |}{[RR^\top]_{jj}} = \left| 1 - 2 \frac{R_{j0}^2}{[RR^\top]_{jj}} \right|,
\end{equation}
and the average concurrence gives a lower bound,
\begin{equation}
	\label{lowerbound}	C_{ave} (\mathcal{Y}) = \sum_{j=1}^m | - R_{j0}^2 + R_{j1}^2 + R_{j2}^2 + R_{j3}^2 | \geq | \sum_{j=1}^m - R_{j0}^2 + R_{j1}^2 + R_{j2}^2 + R_{j3}^2 | =2,
\end{equation}
where the last equality comes from \eqref{orthogonal}.

Let us assume that the lower bound of the average concurrence in \eqref{lowerbound} is attained by an eigenframe with equally distributed concurrence. The concurrence $C$ of each vector must be $1/2$. Let us define $\mathrm{char}_j := 1 - 2 R_{j0}^2/[RR^\top]_{jj}$ so that $C(y_j) = |\mathrm{char}_j|$. We have
\begin{align}
	[RR^\top]_{jj} - 2 R_{j0}^2 &=
	\left\{ \begin{array}{lc}
		\frac{[RR^\top]_{jj}}{2} & (\mathrm{char}_j \geq 0) \\
		- \frac{[RR^\top]_{jj}}{2} & (\mathrm{char}_j < 0)
	\end{array}\right.
\end{align}
Summing both sides of the above equation gives
\begin{equation}
	2 = \sum_{j=1}^m [RR^\top]_{jj} - 2 R_{j0}^2 = 
	\frac{\sum_{\mathrm{char}_j \geq 0}[RR^\top]_{jj} - \sum_{\mathrm{char}_j < 0}[RR^\top]_{jj} }{2}.
\end{equation}
Since $[RR^\top]_{jj}$ are positive and sum to $4$, the right-hand side is equal to $2$ if and only if $\mathrm{char}_j \geq 0$; namely,
\begin{equation}
	1 - 2 \frac{R_{j0}^2}{[RR^\top]_{jj}} = \frac{1}{2}, \qquad j=0,\ldots,n-1.
\end{equation}
Condition \eqref{orthogonal} is necessary and sufficient for an eigenframe to have equally distributed concurrence. Observing that $[RR^\top]_{jj} = \sum_{k=0}^3 R_{jk}^2 = \langle y_j | y_j \rangle$, this condition is equivalent to
\begin{equation}
	\label{ratio}	R_{j0}^2 = \frac{1}{4} \langle y_j | y_j \rangle, \qquad R_{j1}^2 + R_{j2}^2 + R_{j3}^2 = \frac{3}{4} \langle y_j | y_j \rangle,
\end{equation}
which specifies the ratio of the coefficients in the decomposition $\ket[y_j] = \sum_{k=0}^3 R_{jk} \ket[x_k]$.

We are now at the final step in deriving the elegant joint vector from condition \eqref{ratio}. The second equality in \eqref{ratio} reveals that $(R_{j1},~R_{j2},~R_{j3})$ must be on a sphere of radius $\sqrt{3 \langle y_j | y_j \rangle} /2$. These three coefficients are parametrized in spherical coordinates $(\theta_j,\phi_j)$:
\begin{align}
	& R_{j1} = \frac{\sqrt{3 \langle y_j | y_j \rangle}}{2} \sin \theta_j \cos \phi_j, \qquad R_{j2} = \frac{\sqrt{3 \langle y_j | y_j \rangle}}{2} \cos \theta_j,\\
 &\qquad R_{j3} = - \frac{\sqrt{3 \langle y_j | y_j \rangle}}{2} \sin \theta_j \sin \phi_j.
\end{align}
Accordingly, we have
\begin{equation}
	\ket[y_j] = \mathrm{sign} (R_{j0}) \sqrt{\langle y_j | y_j \rangle} \ket[\Psi_{\mathrm{sign} (R_{j0}) \vecn_j}],
\end{equation}
where $\vecn_j = (\sin \theta_j \cos \phi_j,~ \sin \theta_j \sin \phi_j,~ \cos \theta_j )$. Each vector of the eigenframe is of the form given in \eqref{EJV}. The matrix $R$ for the set of vectors $\{ r_j \ket[\Psi_{\vecn_j}] \}_{j=1,\ldots,m}$ is given by
\begin{align}
	\frac{1}{2} \left(\begin{array}{cccc}
		r_1 & \sqrt{3} r_1 n^x_1 & \sqrt{3} r_1 n^z_1 & - \sqrt{3} r_1 n^y_1 \\
		r_2 & \sqrt{3} r_2 n^x_2 & \sqrt{3} r_2 n^z_2 & - \sqrt{3} r_2 n^y_2 \\
		\ddots &&&\\
		r_m & \sqrt{3} r_m n^x_m & \sqrt{3} r_m n^z_m & - \sqrt{3} r_m n^y_m
	\end{array}
	\right).
\end{align}
Condition \eqref{orthogonal} for this matrix $R$ is equivalent to \eqref{zerosum}. When $m=4$, condition \eqref{orthogonal} is equivalent to
\begin{equation}
	\frac{ r_j r_k }{4} (1+ 3 \vecn_j \cdot \vecn_k) = \delta_{jk},
\end{equation}
which indicates that $\vecn_j$ forms a tetrahedron and $r_j = \pm 1$.
$\qed$\\

Theorem \ref{thm:ejm} remains valid even when the functions $C_{max}$ and $C_{ave}$ are substituted with $E_{max}$ and $E_{ave}$, respectively. $E_{max}$ is minimized by exactly the same eigenframes that minimize $C_{max}$, as the former is a bijective and monotonic function of the latter. $E_{ave}$ attains its minimum when $C_{ave}$ does, as $E$ is a convex and monotonic function of $C$.

\end{document}